\documentclass[11pt,aps]{revtex4-1}

\usepackage{graphicx}
\usepackage{natbib}
\usepackage{wrapfig}

\newcommand{\be}{\begin{equation}}
\newcommand{\ee}{\end{equation}}
\newcommand{\nn}{\mbox{} \nonumber \\ \mbox{}}
\newcommand{\ba}{\begin{eqnarray}}
\newcommand{\ea}{\end{eqnarray}}
\newcommand{\om}{\omega}
\newcommand{\Alfven}{Alfv\'{e}n\,}

\newcommand{\E}{{\bf E}}
\newcommand{\B}{{\bf B}}
\newcommand{\J}{{\bf J}}
\newcommand{\A}{{\bf A}}
\renewcommand{\v}{{\bf v}}
\renewcommand{\k}{{\bf k}}
\renewcommand{\div}{{\rm \,div\,}}

\newcommand{\Bf}{{magnetic field}}

\newcommand{\NS}{neutron star}

\newcommand{\NSs}{{neutron stars}}

\newcommand{\Ef}{{electric  field}}

\newcommand{\EM}{{electromagnetic}}

\newcommand\eg{\textit{e.g.,}}

\newcommand\ie{\textit{i.e.}}

\newcommand\lo{\mathrel{\raise.3ex\hbox{$<$}\mkern-14mu\lower0.6ex\hbox{$\sim$}}}
\newcommand\go{\mathrel{\raise.3ex\hbox{$>$}\mkern-14mu\lower0.6ex\hbox{$\sim$}}}

\usepackage{amsmath}
\usepackage[normalem]{ulem}
\usepackage{color}

\begin{document}

\title{Electron MHD: dynamics and turbulence}

\author{Maxim Lyutikov\\
Department of Physics, Purdue University, 
 525 Northwestern Avenue,
West Lafayette, IN
47907-2036 \\
and\\
The Canadian Institute for Theoretical Astrophysics,
University of Toronto, 
60 St. George Street 
Toronto, Ontario, M5S 3H8
Canada}

\begin{abstract}
We  consider  dynamics and turbulent interaction of whistler modes within the framework of  inertialess electron MHD (EMHD). 
We argue there is no energy principle in EMHD: any stationary  closed configuration is neutrally stable. 
On the other hand, the relaxation principle, the long term evolution of  a  weakly dissipative   system towards Taylor-Beltrami  state,  remains valid in EMHD.

We   consider the turbulent cascade of whistler modes.  We show that (i)  harmonic whistlers are exact non-linear solutions; (ii)   co-linear whistlers do not interact (including counter-propagating); 
(iii) waves with the same value of the wave vector, $k_1=k_2$,  do not interact; 
 (iv) whistler modes have  a dispersion that allows a three-wave decay, including into a zero frequency mode; (v) the  three-wave interaction effectively couples modes with highly different wave numbers and propagation angles. In addition, linear interaction of a whistler with a single  zero-mode  can lead to spatially divergent structures via parametric instability.   All these properties are drastically different from MHD, so that  the  qualitative  properties of the \Alfven turbulence  cannot be transferred to the EMHD turbulence.  

We derive the Hamiltonian formulation of EMHD,  and using Bogolyubov transformation reduce  it to the  canonical form; we  calculate the matrix elements for the three-wave interaction of whistlers.   We solve numerically the kinetic equation  and show that, generally, the EMHD cascade develops within  a broad range of angles, while transiently it may show  anisotropic, nearly two dimensional structures. Development of a cascade depends on the forcing (non-universal) and often fails to reach a steady state.  Analytical estimates  predict  the spectrum of magnetic fluctuations for the quasi-isotropic cascade  $\propto k^{-2}$.  The cascade remains weak (not critically-balanced). The   cascade is UV-local,  while the 
infrared locality is weakly (logarithmically) violated.
 
 \end{abstract}

\maketitle

\section{Electron MHD}

In this paper we consider the  plasma dynamics in a specific range of parameters  called  the inertialess electron MHD, or just EMHD \citep{1960RSPTA.252..397L,Kingsep}. 
This limit, by our definition, corresponds to time scales that are much shorter than ion dynamical time scales, but much longer than electron dynamical times scales. Thus,  we  consider the extreme limit, when  the electrons are inertialess, while ions form a neutralizing background. This limit is qualitatively different from what is often called Hall MHD, which  usually means MHD with a Hall term in the Ohm's law. The limit of inertialess electron MHD is, in some sense,  not a smooth transition from MHD, since  inertialess EMHD is not derived from a force balance, but from a constraint condition of zero \Ef\ in the electron fluid frame, see below. 

We have in mind two possible  application of the EMHD cascade (besides the accepted regimes of Solar wind and Earth magnetotail). First, in a weakly collisional astrophysical plasma the MHD cascade couples  to the EMHD cascade at the ion gyro scale \citep{1999ApJ...520..248Q} (see, though Ref. \cite{2013PhRvL.110v5002C}). Secondly, in the \NS\ crusts ions form a neutralizing back ground, while electrons flow slowly, carrying a frozen-in \Bf. Both cases have important   applications, \eg\ for accreting flows around black holes and the behavior of strongly magnetized \NSs, magnetars \citep{TD93,RG}. 

We address a fundamental question: how does an arbitrary  configuration EMHD evolve? This includes the question of stability, the way a perturbed system relaxes and the long term evolution. Typically, it is assumed the the EMHD system behaves similarly to the MHD. In this paper we argue that this is not the case: we demonstrate that any EMHD equilibrium is neutrally stable, so that the system evolves via non-linear wave interaction and forms a turbulent cascade. We then argue that the cascade remains weak (not critically balanced as in case of MHD), and propagates within a broad range of angle (not nearly two-dimensional).

\section{Magnetic field dynamics in electron MHD}


Within the framework of EMHD,  the electric current is  produced by the flow of  electrons, $\J=-ne\v$, where $n$, $-e$, and $\v_e $ are the local number density, charge, and
average velocity of the electrons  \citep[see, \eg\  Ref. ][for derivation and limitations]{Kingsep}. Ions provide a neutralizing background. In the limit of infinite conductivity,   the electric field satisfies the condition,
\be
\E+\v_e \times\B=0,
\label{oo}
\ee
which yields the induction equation
\be
{\partial\B\over\partial t} =  \nabla \times    \left(\v_e  \times \B \right),
\label{induction}
\ee
Replacing
$\J=-ne\v_e ={c\over 4\pi}\nabla\times\B$, the only dynamical variable in this equation is the magnetic field   \citep{Kingsep}:
\be
{\partial\B\over\partial t} = - {c \over 4 \pi  e} \nabla \times    \left({\nabla  \times \B  \over n} \times \B \right)
\label{main}
\ee
 So, given an initial field configuration and appropriate
boundary conditions, the induction equation uniquely determines its evolution.
Importantly, {\it  in the limit of inertialess EMHD, the plasma dynamics follows not from the force balance, but from a constraint condition} (\ref{oo}).

In what follows we consider the basic whistler interactions in the simplest case of constant  rectilinear external \Bf\ and constant density. Dimensionalizing EMHD equations by plasma and cyclotron frequency (so that time is measured in $1/\om_B$ and spacial coordinates in terms of skin depth $\delta = c/\om_p$), absorbing factors of $\sqrt{4 \pi}$ into definition of \EM\ fields,   the  EMHD equation (\ref{main}) becomes.
\be
\partial_t \B =-  \nabla \times (\nabla \times \B \times \B )=  ( (\nabla  \times \B) \cdot \nabla) \B - (  \B \cdot \nabla) \nabla  \times \B={\it {Li}} _\B (\nabla  \times \B)
\label{Li} 
\ee
Thus, the change of the \Bf\ is associated with Lie-transport of the B-vector along the electron fluid with the current, and corresponding  velocity $- \nabla \times \B$.
Appearance of the Lie derivative in the temporal evolution is related to the fact that the EMHD equations  can be written as a Hamiltonian system
 (see Section \ref{Hamiltonian}). Lie derivatives then is equivalent to the Poisson bracket  \citep{1978mmcm.book.....A}.

\subsection{Waves in EMHD: fully non-linear whistlers}
\label{non-linear}

Let us assume that constant   \Bf\ $\B_0=B_0  {\bf e} _x  $  is directed along $x$ axis, and the wave is propagating at angle $\theta$ with respect to the \Bf.  First,  note that a   harmonic wave of {\it  arbitrary}  amplitude $ \delta B$  is a an exact solution of EMHD 
\ba && 
\B=B_0  {\bf e} _x+ \delta B {\bf e}_B e^{-i (\om t - k (x \cos \theta + z \sin \theta))}
\nn &&
\om =   \B_0  k^2 |\cos \theta| 
\nn && 
{\bf e}_B={1\over 2} \,  \{ - i  \sin \theta, 1 , i \cos \theta \} 
\label{whistlers}
\ea
where  ${\bf e}_B$ is  the eigenvector (regaining the constants, $\om = c^2 k^2 | \cos \theta |  {\om_B / \om_p^2} $). 

The whistler modes parametrized by (\ref{whistlers}) allow a universal treatment of the  field-alined and  counter-aligned propagation, without a need to introduce separate polarization states \citep[as, \eg\  done in Ref. ][]{2003PhPl...10.3065G}.

The fact that an arbitrary amplitude harmonic wave is a solution of EMHD equation {\it does not mean} that an  wave cannot experience a decay. Recall, in MHD an arbitrary  \Alfven wave, not necessarily harmonic, is an exact solution for parallel propagation. In the case of EMHD, only harmonic waves are  non-linear solutions. Two non-collinear  harmonic waves will produce a third, beat wave. Or, reversely, a given wave can decay in two.

There are  special cases when the strong whistler  waves do not interact. Let as assume that there two fully non-linear modes  with wave numbers $k_1$ and $k_2$ propagating  at angles $\theta_1$ and $\theta_2$ (not necessarily co-planar). One can demonstrate by direct calculations that 
 two fully  non-linear harmonic waves are completely non-interacting for (i) co-linear propagation (including counter-propagating case); (ii) when the  two modes have the same wave-vector $k_1 =k_2$ (but  propagating in different directions), see below Section  \ref{inter}

An initial perturbation with a broad range of wave vectors can be represented as a sum of harmonic waves. For aligned propagation (so that the  initial perturbation is not experiencing angular spreading) these  waves are non-interacting, but since different modes have different velocities and initial perturbation will experience a dispersive spreading along the direction of propagation. 


\section{Conservation laws}

Equations of EMHD conserve energy
\ba &&
\partial_t B^2/2 + \nabla \cdot( \E \times \B)=0
\nn &&
\E = \nabla \times \B \times \B= \J \times \B
\nn &&
\E \times \B  =   (\nabla \times \B \cdot \B ) \B - B^2 \nabla \times \B
\label{EEE}
\ea 
Note that  the two terms in the Poynting flux separate energy transport by velocity $v= - \nabla \times \B$ and parallel  energy transport.
The current flow is divergenceless as well:
\be 
\J = \nabla  \times \B
, \,
\nabla \cdot \J =0
\ee 
The equations of EMHD cannot be written in fully conservative form (as a divergence of the stress-energy tensor), since the force balance in not included in the EMHD formulation: there  are tensile forces that balance Lorentz force $\J\times \B$ , but those forces do no work,  $\v\cdot \J\times \B=0$, so that energy is conserved.

In EMHD the helicity is conserved:
 \ba &&
  \partial_t  (\A \cdot\B) = (\B \cdot  \nabla) f - \A \cdot \nabla \times ( \J \times \B) = (\B \cdot \nabla) f + \nabla \cdot ( \A \times (  \J \times \B))
\nn &&
\int dV  \partial_t  (\A \cdot\B) = \int dV (\B \cdot\nabla) f + \int d {\bf S} \cdot( \A \times ( \J \times \B))
\ea

\be
\int dV (\B \cdot\nabla) f  =\int dV ( \nabla ( f \B) - f \nabla \B) = 
\int  f \B \cdot d {\bf S} =0
\ee
So a change of helicity in a volume is due to flux of $- \A \times  \J \times \B$ through the surface. It is zero if ${\bf n} \cdot \B = {\bf n} \cdot \J=0$, no \Bf\ or current penetration through the surface.

In a  differential form the conservation of helicity can be written as a  conservation of the topological torsion four-vector.
\be
\partial_t (\A \cdot \B)= - \nabla (\A \times \dot{\A}) 
\ee

Though the  condition ${\bf n} \cdot \J=0 $ on the boundary  seems to be   an additional constraint in EMHD if compared with MHD, in fact, it is not: just like in ideal MHD the boundary must move with the flow, in ideal EMHD the boundary must move with the current, so no current  actually crosses the boundary.

\section{The energy and the relaxation principles in  EMHD}
\label{123}

In MHD the energy principle \citep{1958RSPSA.244...17B} and the relaxation principle  \citep{Woltjer58,Tayler73}  play  the most important role, allowing one to decide on a stability of a plasma  configuration and its long term evolution  without doing a full normal mode analysis. The energy principle allows one to decide on the stability of a particular stationary configuration by calculating an energy change induced by small perturbations: if the energy increase, then the configuration is linearly  stable. The  relaxation principle  addresses the long term evolution of a system. In a viscose but electrically ideal MHD system,   by  dissipating the energy yet conserving the topology, the plasma  tries to relax to the maximally twisted state, the  force-free Woltjer-Taylor state. Somewhat similarly, in a weakly resistive medium magnetic energy cascades to smaller scales where dissipation is important, while helicity concentrates on larger scales and is thus better conserved.

As we show below, the case of EMHD is qualitatively different from MHD: there is no energy principle in EMHD. Recall,  in MHD any stationary {\it closed}   state
 is the result of the force balance. Conservative forces can be expressed as gradients of the corresponding  potentials. Then  the stable equilibria correspond to minima of the potential energy. The crucial ingredient in the energy principle is the fact that {\it the energy of a closed system can change due to internal motions}. Indeed,  in MHD a work of  Lorentz force is generally not zero, 
$
\int dV \v \cdot  \J \times \B \neq 0.
$
This is {\it not}  true in EMHD: the work of the  Lorentz force is identically zero, 
\be
 \v \cdot  \J \times \B \propto \J \cdot \J \times \B =0.
 \ee
 Thus,  the energy of a closed  EMHD plasma cannot change. This immediately implies the absence of the energy principle, which {\it requires} that the magnetic energy  tries to reach a minimal energy state. 
As another way to understand the absence of the  relaxation principle in EMHD, recall that the equations of ideal  EMHD   are derived from the constant equation (that the \Ef\ in the frame of the electron fluid is zero), and {\it not} from the force balance equation, like, \eg\ the MHD equations. 


\subsection{Any closed EMHD equilibrium is neutrally stable}
Mathematically, to prove the absence of the energy principle, let us prove that {\it any} stable configuration of EMHD plasma with constant plasma density has purely real frequency: it is neutrally stable. Let $B_0$ be a stationary solution of Eq. (\ref{Li}), $\partial_t \B_0=0$. 
Let us consider an incompressible  electron fluid perturbation $\xi({\bf r})$, so that the perturbed \Bf\ is
\be
\delta\B = \nabla \times(\xi \times \B_0) =  {{\cal{ L} }i } _{\B_0} \xi
\ee
We find
\ba && 
\dot{\xi} \times \B_0+\delta {\bf F} =0
\nn &&
 \delta {\bf F} = \J_0 \times \delta \B + \delta \J \times B_0 
 \ea
 It is well known that operator $\delta {\bf F} (\xi)$ is self-adjoint \citep{Kulsrud}.
 For harmonic time variation $\propto e^{- i \om t}$ we find 
 \be
\int dV   \eta^\ast \cdot \dot{\xi} \times \B_0= - i \om  \int dV  \eta^\ast \cdot {\xi} \times \B_0 = i \om  \int dV  \xi \cdot \eta^\ast  \times B_0
\label{1}
 \ee
 On the other hand, using self-adjointness of ${\bf F} $:
  \be
\int dV   \eta^\ast \cdot \dot{\xi} \times \B_0=-   \int dV   \eta^\ast \cdot  {\bf F} (\xi)=-  \int dV \xi \cdot  {\bf F} (\eta^\ast ) = \int dV \xi \cdot   \dot{\eta^\ast } \times \B_0 = 
i \om^\ast  \int dV \xi \cdot  {\eta^\ast } \times \B_0
\label{2}
 \ee
 Comparing Eqns (\ref{1}) and   (\ref{2}) we conclude that $\om = \om^\ast$, that is, $\om$ is  purely real.
This demonstrates that {\it any}  stationary configuration of EMHD is neutrally stable to linear perturbations.

In fact, one can demonstrate the non-linear stability of  an arbitrary EMHD  equilibrium, not necessarily a closed one. By the Poynting theorem, Eq.  (\ref{EEE}), 
the energy within a volume can only change due to flow of energy through the boundary. (Again, this is qualitatively different from MHD, where \Bf\ energy can change due to internal motion within a confined volume.) Consider harmonic perturbations of a given EMHD  equilibrium $\B_0$. Generally, there will be two types of perturbations: those that do excite surface motion and those the do not.  (Note, that the electron fluid element displacements at the star surface does not imply crustal motions.)  If the outside medium is vacuum, the perturbations that do excite harmonic in time surface motion will be radiated away as \EM\ Poynting flux (with local  intensity $B_n^2 \xi^2 \om^2 /c$ where $\xi$ is the amplitude of the electron fluid motion and $B_n$ is the normal component of \Bf).  Perturbations that do not excite surface electron motion remain neutrally stable. 
Thus, {\it any} EMHD configuration is neutrally stable.
Neutrally stability implies that none of the stable configurations work as attractors.

The above result, \ie,  the absence of the energy principle in EMHD,  demonstrates that conditions for MHD stability cannot be directly applied to the EMHD stability, and many fundamental results has to be abandoned. For example, it is well known that in MHD purely toroidal or purely poloidal configurations are unstable. This is not true in EMHD: there can be neutrally stable purely toroidal and purely poloidal configurations!

Let us summarize.  An EMHD stable state is qualitatively different from the MHD stable state. If a stable MHD  state is perturbed, then according to the energy principle the new state has a higher energy, so a system tries to relax to the preferred stable state. If an EMHD stable state is perturbed, the system does not relax back to this state

\subsection{Relaxation principle}

The energy principle  discussed above addresses the short time scale evolution of a system. On the other hand, in MHD a different principle, the relaxation principle  \citep{Woltjer58,Tayler73}, 
addresses the long term evolution of the MHD system. The relaxation principle tries to predict, without detailed calculations the long term evolution of a system.  Topological constraints, like conservation of helicity, restrict a set of final states.
 The  relaxation principle of MHD \citep{Woltjer58,Tayler73}  then states that an arbitrary MHD state first relaxes to a  minimal energy state that has the same helicity as the initial state. In electrically ideal plasma  the helicity is conserved exactly, while the energy can be dissipated via fluid viscose effects. In a weakly resistive plasma, helicity is concentrated on largest scales and thus is dissipated slower than the magnetic energy. 
 
 Since EMHD conserves helicity, the Taylor relaxation principle remains valid in EMHD as well: if an EMHD state is perturbed, the energy will cascade down to small scales, while helicity will be concentrated on largest scales. But the largest scales will have very little energy. In reality, the pure Taylor-Beltrami state \cite{Lamb} is not achievable due to realistic  boundary conditions.

\section{MHD   and  EMHD turbulence: the key differences}

After the initial ideas on the MHD turbulence were laid out 
in Ref. \cite{1963AZh....40..742I} and \cite{1965PhFl....8.1385K}, 
the modern theory of MHD turbulence is based on the Goldreich-Sridhar model \citep{1994ApJ...432..612S,1995ApJ...438..763G,1997ApJ...485..680G}, whereby 
the turbulence is described in terms of interacting, counter-propagating wave packets. The four-wave processes dominate, but three wave process, when one of the mode has zero frequency, also contribute to the cascade \citep{1983JPlPh..29..525S,1995ApJ...447..706M,1996ApJ...465..845N,1997ApJ...485..680G}. Interaction is local in $k$-space (so that only waves with similar wave numbers interact), leading to Kolmogorov-like cascade. The cascade is anisotropic \citep[possibly even 3D anisotropic,][]{2006PhRvL..96k5002B}, but the parallel and perpendicular scales are related to each other via the critical balance assumption.

Currently, the dominant view is that the development of the EMHD turbulence proceeds  similarly to the MHD cascade, \ie, that the turbulence becomes strong and as a result achieves a critical balanced state. 
The scaling of EMHD cascade was proposed to be either $7/3$ \citep{1973JETP...37...73V,1999PhPl....6..751B,2003PhPl...10.3065G,2004ApJ...615L..41C,  2008PhRvL.100f5004H} or $8/3$ \citep{2012ApJ...758L..44B}. 
The origin of these models, and the $7/3$ scaling can be traced back to the Ref.  \cite{1973JETP...37...73V}. Briefly, the frequency of vortex breaking at scale $1/k$ is $\nu_k \sim B_k k^2$ (from the EMHD equation (\ref{main})); assumption of constant energy flux in $k$ space implies 
$B^2_k \nu_k =$const, so that $B_ k \propto k^{-2/3}$. The spectrum is then $E_k \sim B_k^2 /k \propto k^{-7/3}$.  

 The work in Ref.  \cite{1973JETP...37...73V}  was criticized in Refs. \cite{Kingsep,RG}, where it was argued that the EMHD turbulence always remains weak. (This criticism also applies to later works that used the scaling of Ref.  \cite{1973JETP...37...73V}, \citep{1999PhPl....6..751B}  and others).  Rephrasing the works \cite{Kingsep,RG},   in Ref. \cite{1973JETP...37...73V}  it was assumed that the whistler mode interaction is based on a concept of a vortex, a long-lived coherent structure. Due to highly dispersive nature of the whistler modes, an initial  packet with a spread $\Delta k\sim k $, will have phases mixed on time scale 
 $t_{mix} \sim 1/( k v_p) \sim 1/\om$. That is, a coherent vortex,  or a wave package,  falls apart on its own on its dynamical time without transferring  energy to smaller scale, so that a concept of a vortex is inconsistent with its  suggested life time. Thus, 
 the period of whistlers at scale  $k$ is smaller that the inferred  life time of a vortex   by $B_k/B_0$. The non-linear interaction time is then longer by $B_0/B_k$; the constancy of energy flux in the $k$ space then implies $B_k \propto k^{-1/2}$, $E_k  \propto k^{-2}$ \citep{RG}. This implies that EMHD cascade proceeds in the weak regime. Our numerical  calculations, generally,  confirm the picture suggested in Ref.   \cite{Kingsep,RG}: $E_k  \propto k^{-2}$ spectrum follows from the form of the matrix element, while the resulting   cascade remains weak.
 
Another important issue related to the structure of the cascade is its anisotropy. Below in Section \ref{Solving} we show that often 
 the whistler cascade becomes quasi-isotropic, while  generically highly anisotropic structures can appear transiently.

\section{Hamiltonian formulation of EMHD and the three-wave  interaction matrix}
  \label{Hamiltonian}
  
Turbulent processes are best described in the Hamiltonian formulation  \citep{1997PhyU...40.1087Z}. In this section we derive the  Hamiltonian formulation of EMHD and use it to derive the EMHD  three-wave  interaction matrix. Note, that in the conventional Hamiltonian formulation of MHD the  velocity (or velocity potential) is an independent variable  \citep{1997PhyU...40.1087Z}.  In contrast, in EMHD 
velocity is not an independent variable; this requires a different Hamiltonian formation,  a different set of canonical variable.

Non-linear wave-wave interaction is usually considered in the occupation representation of mode excitation \citep{Zakharov,1997PhyU...40.1087Z}. As an elementary mode we take 
\ba && 
\delta \B = {b \sqrt{\om}}  {\bf e}_B e^{- i \Phi}
\nn &&
  {\bf e}_B={1\over 2} \,  \{ - i  \sin \theta, \cos \phi + i \cos \theta \sin \phi , i \cos \theta \cos \phi - \sin \phi \} 
  \nn &&
  \Phi = \om t  -\k  \cdot  {\bf r}
  \label{OBB}
\ea
where $b$ is the mode amplitude, which corresponds to the   creation  operator in quantum mechanics. The angular variables are spherical coordinates of the wave vector in the frame aligned with \Bf.
This choice of variables $b$  effectively accomplishers the Bogolyubov transformation \citep{Bogolyubov}, so that in the absence of interactions 
\be
\partial_t b(\k, t) + i \om  b(\k, t)=0, 
\ee
 while the energy of the waves takes the canonical form
\be
{\cal H} = \sum_k \om_k b_k b^\ast_k
\ee
In  these variables the linearized equations of motion become trivial, so that all the linear dynamic is incorporated into the dispersion relation $\om(\k)$.

The Lagrangian of EMHD includes the \Bf\ energy density and a constraint equation  (\ref{Li}). Let us chose $\B$ as an independent coordinate (another choice is the vector potential); adding a constraint with  a Lagrangian multiplier, the EMHD  Lagrangian is 
\be
 {\cal L} = - B^2/2 + {\bf S} \cdot(  \dot{\B} + \nabla \times (  \J \times \B) )
 \ee
Requirement of the stationary action is expressed in terms of the generalized Euler-Lagrange equation \cite{1996qft..book.....R}
\be
{\delta {\cal L} \over \delta B_i} - \partial_t \left( {\delta {\cal L} \over \delta \dot{ B}_i} \right) - \nabla_k   \left( {\delta {\cal L} \over \delta{ B}_{i,k}} \right)=0.
\ee
This gives   the dynamical equation for the Lagrange multiplier ${\bf S}$, which becomes the canonically  conjugate variable to $\B$.
\ba &&
{\delta {\cal L} \over \delta B_i}= -\B + ( \nabla \times    {\bf S} \times  \nabla \times  \B) 
\nn &&
 \partial_t \left( {\delta {\cal L} \over \delta \dot{ B}_i} \right) = \dot{  {\bf S} }
 \nn &&
  \nabla_k   \left( {\delta {\cal L} \over \delta{ B}_{i,k}} \right)=  \nabla \times (\B \times  \nabla \times    {\bf S}) 
  \ea
  \be
  \dot{  {\bf S} }=-\B +  ( \nabla \times    {\bf S} \times  \nabla \times  \B)  + \nabla \times (\B \times  \nabla \times    {\bf S}) 
  \ee
  The Hamiltonian is then
   \be
 {\cal H}= {\bf S} \cdot \dot{\B}-  {\cal L}=B^2/2 - \left( \nabla \times   ( \nabla \times   {\bf B} \times   \B)  \right) \cdot   {\bf S} 
\ee
 
 Following our observation  above, Eq. (\ref{OBB}), we first do the Fourier transform, then
 canonical transformations $\B \rightarrow \sqrt{\om} {\bf B}$, ${\bf S}  \rightarrow  {\bf S}/ \sqrt{\om}$, and then 
change to complex variables
\ba &&
{\bf b} = (\B + i    {\bf S}) /\sqrt{2}, \, {\bf b} ^\ast =  (\B - i    {\bf S}) /\sqrt{2}
\nn &&
\B= { {\bf b}  + {\bf b} ^\ast \over \sqrt{2}} , \,     {\bf S}= i { {\bf b} ^\ast - {\bf b}  \over \sqrt{2}}
\ea
with the equations of motion
\ba && 
\partial_t {\bf b} = - i  {\delta {\cal H} \over \delta {\bf b}^\ast} 
\nn &&
\partial_t {\bf b}^\ast = i   {\delta {\cal H} \over \delta {\bf b}}
\ea
In the new variables the  Hamiltonian takes the  form
\ba &&
{\cal H} = {\cal H}_0 +{\cal H}_{\rm int}
\nn &&
{\cal H}_0 = \om_k b_k b_k ^\ast
\nn &&
{\cal H}_{int}  = \int d\k_1 d\k_2  \left( 
 {1\over 2}  V_{k\leftrightarrow 1,2} b^\ast  b_1 b_2 \delta(\k  - \k_1 -\k_2)    + V^\ast_{ 1\leftrightarrow k,2}b  b_1 b_2^\ast  \delta(\k_1 - \k  -\k_2) \right) 
   \nn &&
    V_{k\rightarrow  1,2} = \sqrt{ \om_1 \om_2 \over  \om} \left\{  {\bf e}_B^\ast \cdot \left[\left((\k_1 \times {\bf e}_{B_1}) \cdot \k_2 \right) {\bf e}_{B_2} - (\k_2 \cdot {\bf e}_{B_1}) (\k_2 \times {\bf e}_{B_2})   + (1 \leftrightarrow 2)  \right] \right\}
\ea

\section{The three-wave resonant condition for whistlers}
\label{three-wave}

Two ingredients are important for the wave-wave interaction: the resonant condition that relates kinematically the properties of the participating waves and the interaction matrix, that determines the strength of the interaction. Let us first  study the resonant condition.

The normal modes of EMHD, whistlers, have a
 dispersion relation of the decaying type, which resembles shallow water waves. Thus, in the weak turbulence regime  we expect that 3-wave processes are the dominant.
The three-wave resonant condition $\delta (\k_0  - \k_1 -\k_2) \delta ( \om_0  - \om_1 - \om_2)$ for whistler waves reads
\ba && 
\om_n = k_n ^2 |\cos \theta_n|, \, n=0,1,2
\nn && 
\om = \om_1 + \om_2
\nn &&
k \cos \theta =k_1 \cos \theta_1+  k_2 \cos \theta_2 
\nn &&
k \sin \theta  \sin\phi  = k_ 1 \sin \theta_1 \sin\phi_1 + k_ 2 \sin \theta_2 \sin\phi_2
\nn &&
k \sin \theta  \cos\phi  = k_ 1 \sin \theta_1 \cos\phi_1 + k_ 2 \sin \theta_2 \cos\phi_2
\label{res}
\ea
(below we drop subscript $0$ on the parent wave, $k\equiv k_0$).

The resonant conditions can be resolved for $\{ k , \theta, \phi \}:$
\ba && 
\cos \theta  = {( k_1 \cos \theta_1 + k_2 \cos \theta_2)^2 \over  k_1^2 | \cos \theta_1| + k_2 ^2 |\cos \theta_2|}
\nn &&
k  = {  k_1^2 | \cos \theta_1| + k_2 ^2 |\cos \theta_2| \over k_1 \cos \theta_1 + k_2 \cos \theta_2}
\nn &&
\tan \phi = {k_1 \sin\theta_1 \sin \phi_1 + k_2 \sin \theta_2 \sin \phi_2 \over k_1 \sin\theta_1 \cos \phi_1 + k_2 \sin \theta_2 \cos \phi_2}
\label{kkkq}
\ea


\subsection{Co-planar wave interaction}
\label{Co-pla}

General resonance condition (\ref{res}-\ref{kkkq}) is complicated; let us gain some insight into  the non-linear wave interaction by  considering  a special cases,  a decay (and a merger of) into co-planar (with \Bf) propagating modes. As we discuss below, even this simple case is very complicated. First we consider wave decay,  $k;1,2$ processes (the semicolon separates the parent wave from the daughter waves). Later, in Section \ref{conf}, we consider the resonant conditions for wave confluence  $k;1,2$.

\subsubsection{Co-planar decay of obliquely propagating waves}

\begin{figure}[h!]
\includegraphics[width=.49\linewidth]{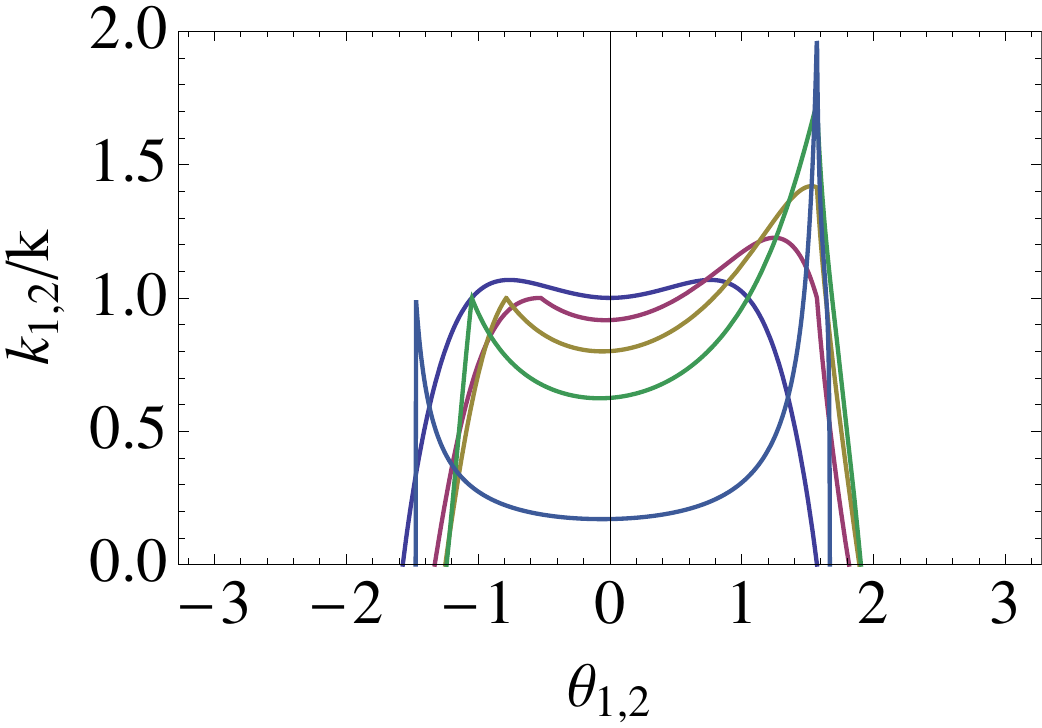}
\includegraphics[width=.49\linewidth]{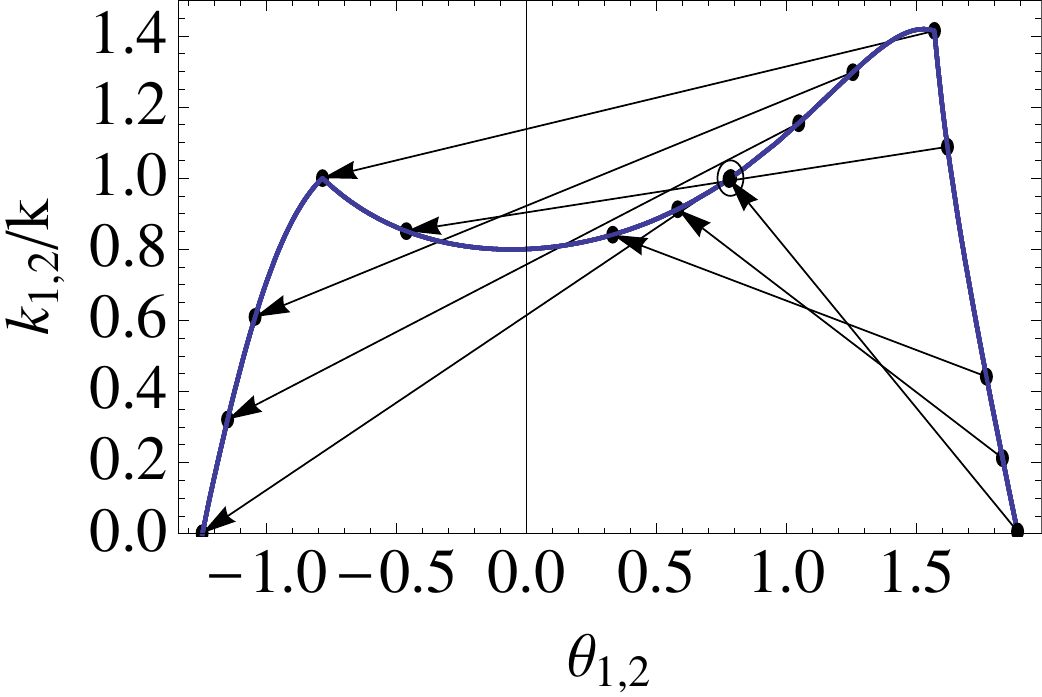}
\caption{{\it Left Panel}. Dispersion relation for the decay modes $\{k_1, \theta_1\}, \{k_2, \theta_2\}$,  oblique propagation  of the parent-wave, $\theta= 0, \, \pi/6,  \, \pi/4, \, \pi/3, \, \pi/2-0.1$ (becoming more peaked at $\theta_{1,2} = \pi/2$ for higher initial angles). The maximal wave number for  daughter wave is $2 k$, reached for a highly oblique parent mode.  Negative values of $\theta$ indicate emission of a wave propagating in the direction  opposite to the parent mode  along the $y$ axis. {\it Right Panel}. Paring of the two daughter modes, $\theta = \pi/4$.   }
\label{Oblique-decay}
\end{figure}

Setting $\phi_i =0$ and eliminating $k_1, k_2$  from (\ref{res}), one can derive the following relation between the angles of the waves: 
\be
|\cos \theta| \sin^2 (\theta_1-\theta_2) = |\cos \theta_1| \sin^2 (\theta-\theta_2) + |\cos \theta_2| \sin^2 (\theta-\theta_2) 
\ee
Alternatively, eliminating $\{k_2, \theta_2\}$ from the momentum conservation,
\ba && 
k_2 = \sqrt{k^2 +k_1^2 - 2 k k_1 \cos(\theta-\theta_1)}
\nn &&
\cos\theta_2= \pm { k \cos \theta -k_1 \cos \theta_1 \over \sqrt{k^2 +k_1^2 - 2 k k_1 \cos(\theta-\theta_1)}},
\ea
 one can derive the following resonance condition
\be
k^2 \cos \theta = k_1^2 |\cos \theta_1| +
\sqrt{k^2 +k_1^2 +2 k k_1 \cos(\theta-\theta_1)} | k \cos \theta - k_1 \cos\theta_1|
\label{kjk}
\ee
For given properties of the parent wave $\{k, \theta\}$,  Eq. (\ref{kjk}) is an equation for $k_1 (\theta_1)$ (it is also an equation for $k_2 (\theta_2)$ after a substitution $k_1 \rightarrow k_2$ ,
$ \theta_1\rightarrow  \theta_2$.) Its solution is 
\ba&&
k_1 = \left( {\cos ^2  \theta + 2 \cos  \theta \cos  \theta_1+ 4 \cos  \theta \cos( \theta- \theta_1) \cos  \theta_1 + \cos^2  \theta_1} \pm 
\right.
\nn &&
\left. {
(\cos  \theta-\cos  \theta_1) \sqrt{ \cos^2  \theta + 2 \cos  \theta(3-4 \cos( \theta- \theta_1) \cos  \theta_1 +\cos^2  \theta_1}}
\right)  { k\over 2 (3 \cos \theta + \cos (\theta-2 \theta_1)) \cos \theta_1} ,
\label{kjk1}
\ea
see Fig. (\ref{Oblique-decay}).
For a given $\{k, \theta\}$ the angle $ \theta_1$ can be chosen to lie in the range $-\pi/2 <  \theta_1 <  \pi/2$. The resonance condition will then give $\{k_2, \theta_2\}$ that generally may be propagating opposite to the initial wave, $\theta_2 > \pi/2$.
For a simple decay, the wave vector of the daughter mode  $k_1$ can be up to two times larger than $k$, $k_1 \rightarrow  2 k$ for  $\theta, \, \theta_1 \rightarrow \pi/2$.  

\begin{figure}[h!]
\includegraphics[width=.49\linewidth]{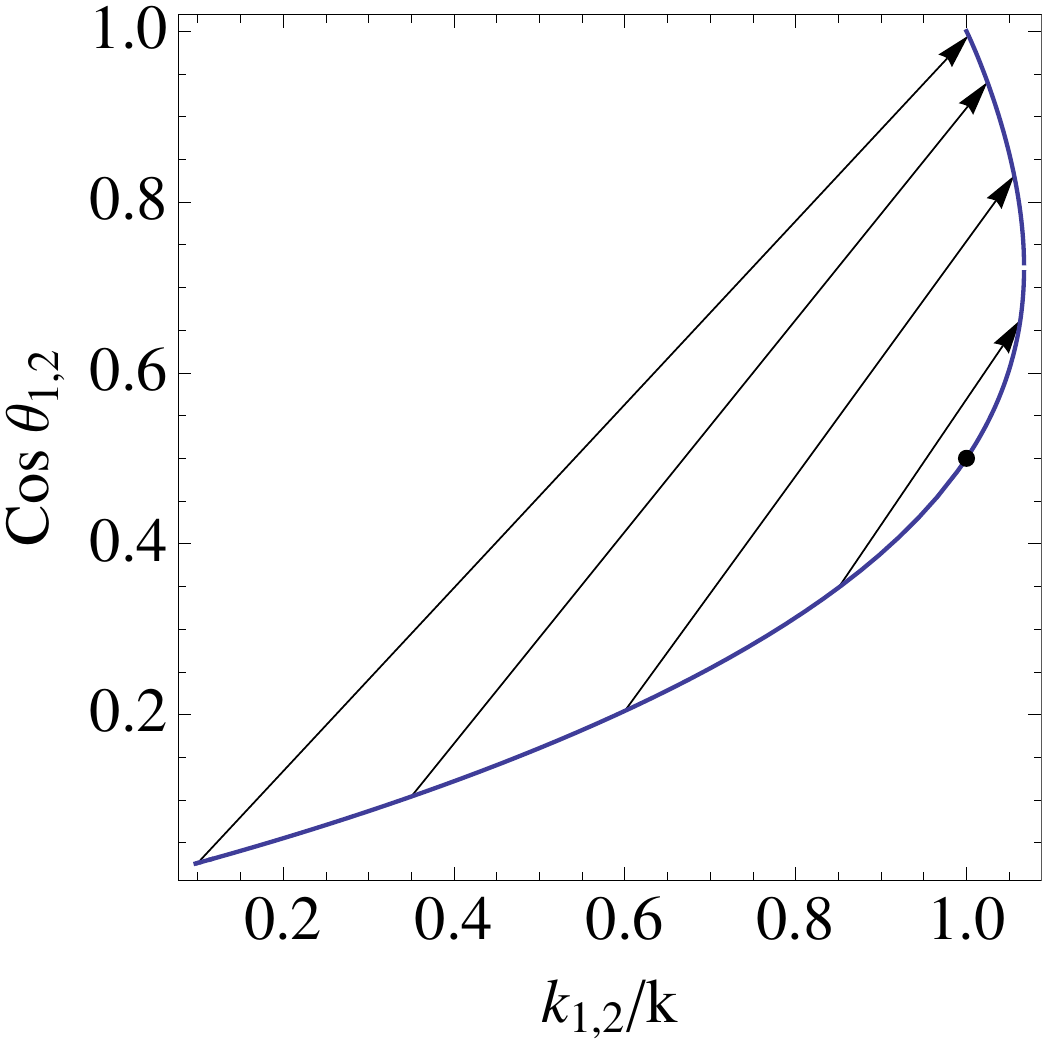}
\includegraphics[width=.49\linewidth]{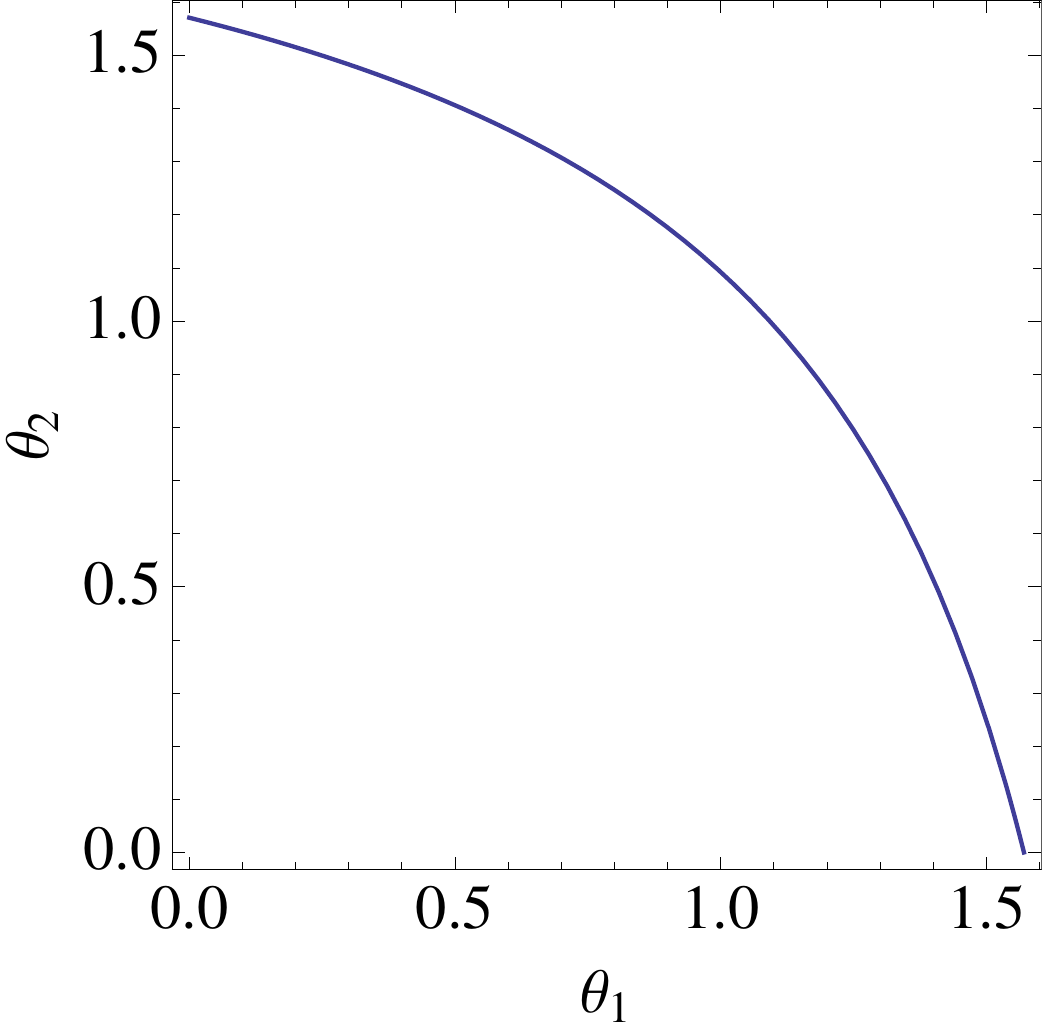}
\caption{{\it Left Panel}. Dispersion relation for the decay modes $\{k_1, \theta_1\}, \{k_2, \theta_2\}$, parallel propagation of the parent-wave. Arrows connect the  two decay product  waves.  The small dot separates the two product waves. Two daughter modes  propagate in the opposite direction along $y$ axis. {\it Right Panel}. Relation between the angles of the two daughter modes for a decay of a parallel-propagating initial mode (absolute values of $\theta_1$ and $\theta_2$ are plotted).}
\label{Res=parallel}
\end{figure}

General relations take especially simple form for the parent wave propagating along \Bf:
\be
k_{1,2} = { 1+2 \cos  \theta_{1,2}+ 5 \cos ^2  \theta_{1,2} + (1-\cos  \theta_{1,2})^{3/2} \sqrt{1+ 7 \cos  \theta_{1,2}}\over  4 \cos  \theta_{1,2} (1+ \cos  ^2 \theta_{1,2})} \, k,
 \ee
 see Fig \ref{Res=parallel}. The dispersions curve in Fig. \ref{Res=parallel} implies that a whistler mode propagating along \Bf\ can decay  only into two forward propagating waves.
For parallel propagating parent wave  the angles of the two daughter modes are related via
\be
 \cos \theta_2 ={1+(2-3  \cos  \theta_1)  \cos  \theta_1+ \sqrt{ 1-\cos  \theta_1} (1+ \cos  \theta_1) \sqrt{1+ 7  \cos  \theta_1} \over 2+ 6 \cos  \theta_1+  \cos  ^2\theta_1},
\ee
see Fig. \ref{Res=parallel}.

One of the daughter modes has $\{k_1<k, \, \theta_1> \pi/3\}$, another  $\{k_2> k, \, \theta_2< \pi/3\}$. The maximal wave number is reached at $\theta_2= 0.76$ and is equal to $k_2 = 1.06 k$. 
Note, that for the inverse process,  a merger of two waves, {\it  in order to create a whistler propagating along \Bf\ with wave number $k$, one of the waves should have $k_1 >k$. }
As a special case,  there is  kinematically allowed  decay of $\{k,\theta=0\}$ mode into two modes  $\{k,\theta=\pi/3\}$, (daughter modes have  the same $k$ as the initial wave). As we discuss  in Section \ref{non-linear}, such process actually does not occur (zero matrix element).

\begin{figure}[h!]
\includegraphics[width=.99\linewidth]{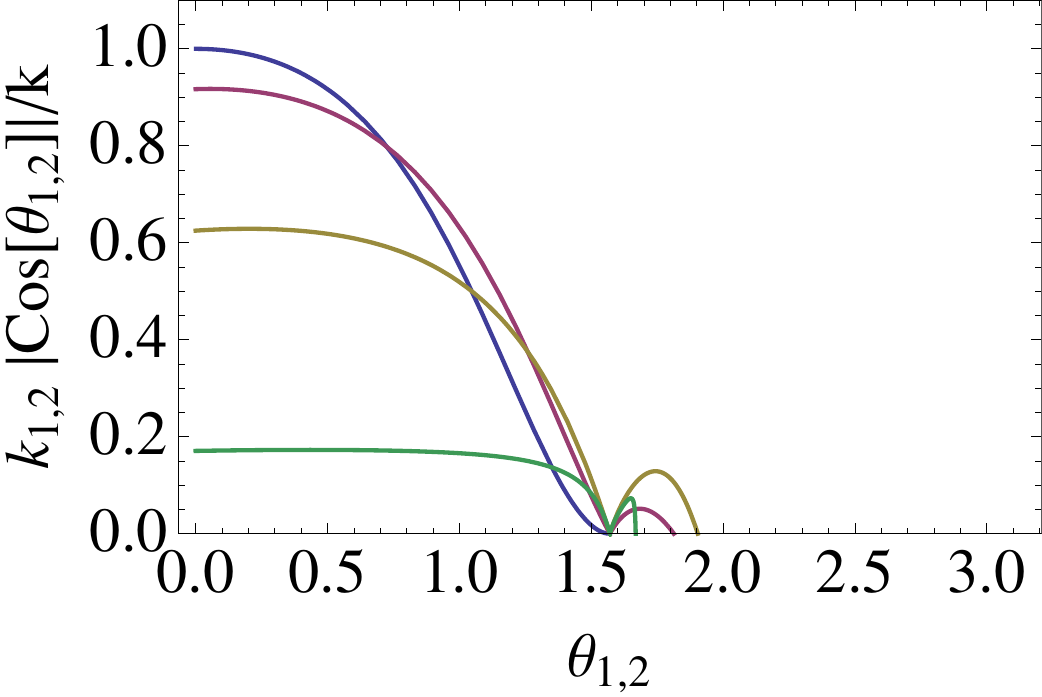}
\caption{Parallel wave vector for the decay modes (see Fig. \ref{Oblique-decay} for notations). All daughter modes have  $k_{\parallel; 1,2} < k$. }
\label{Oblique-decay-kpara}
\end{figure}

Let us highlight a number of important features. (i) If the parent mode propagates at angle $\theta$, one of the daughter modes has $\theta_1 > \theta$. For  $\theta< \theta_1 < \pi/2$, the other wave has $-\pi/2 < \theta_2 < - \theta$.  (ii) obliquely propagating mode can decay into backward moving wave,  $\theta_1 > \pi/2$. The maximal angle for a backward propagating wave is $\theta_{1,max} = 
\pi /2 + \arccos (4 \sqrt{3}/7)$;  it is reached for $\theta =\pi/3$, in this case $\theta_2 (\theta_{1,max}, \theta=\pi/3)=0$. For $\theta > \pi/3$ and $\theta_1 > \pi/2$, both product waves propagate in the same direction along $y$ axis, $\theta_2 > 0$. As $\theta_2 \rightarrow \theta-0$, $\theta_1\rightarrow \pi/2+0$. (iii) All the daughter modes have  $k_{\parallel; 1,2} < k_\parallel$, Fig. \ref{Oblique-decay-kpara}.

\begin{figure}[h!]
\includegraphics[width=.49\linewidth]{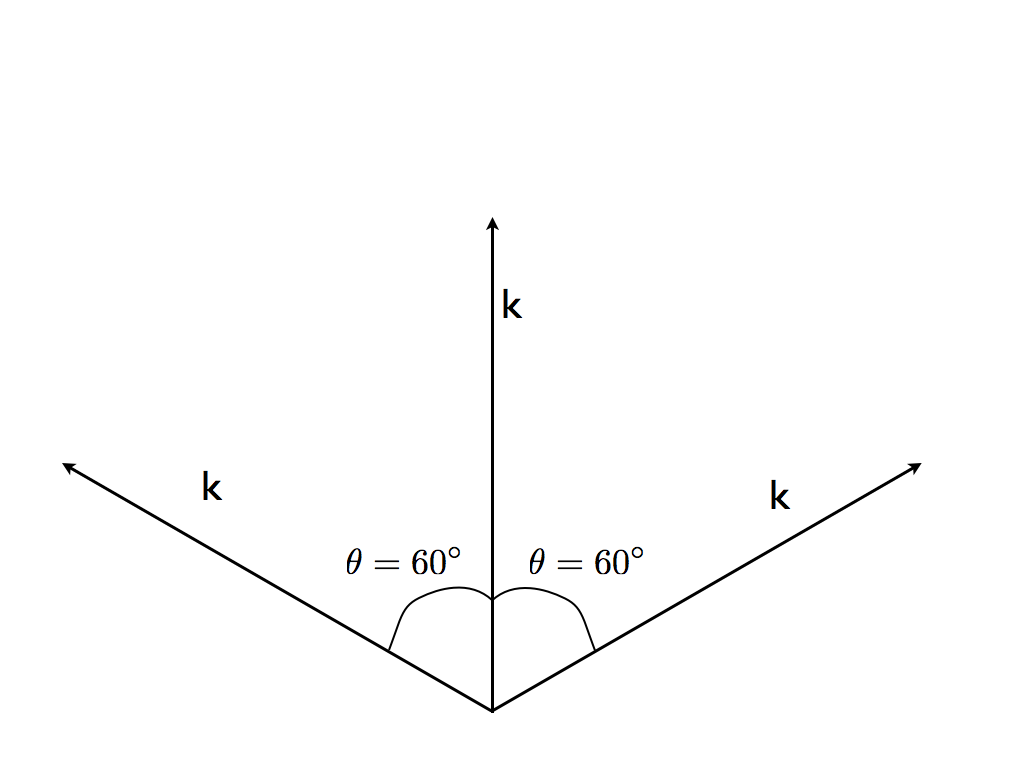}
\includegraphics[width=.49\linewidth]{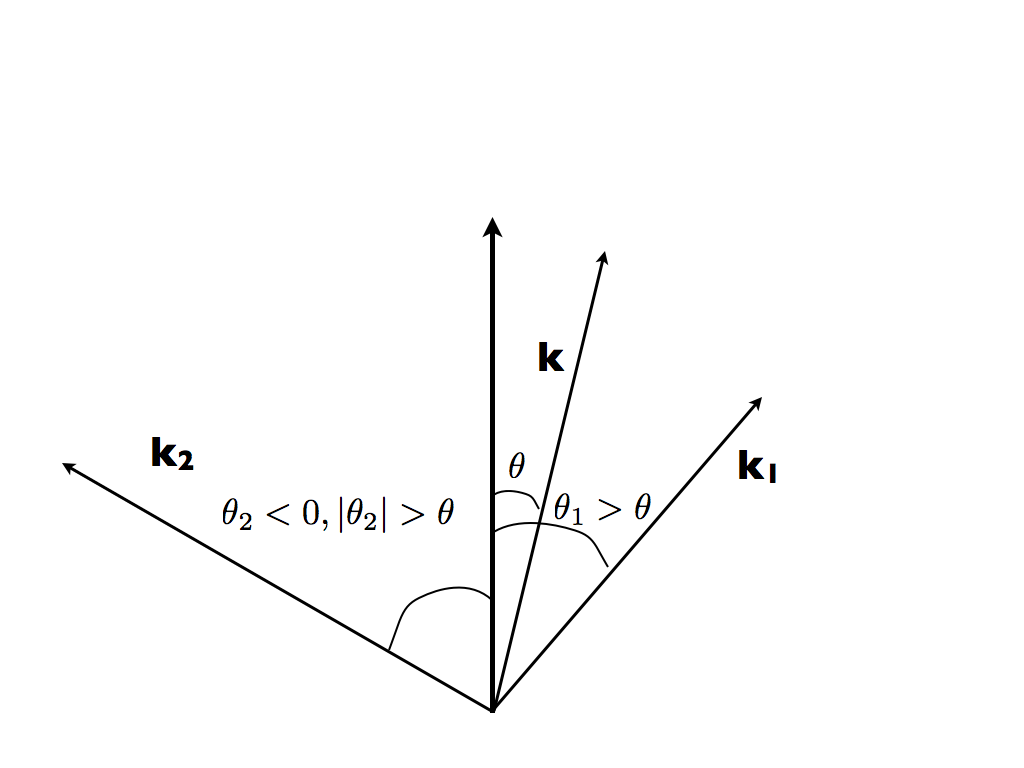}
\\
\includegraphics[width=.99\linewidth]{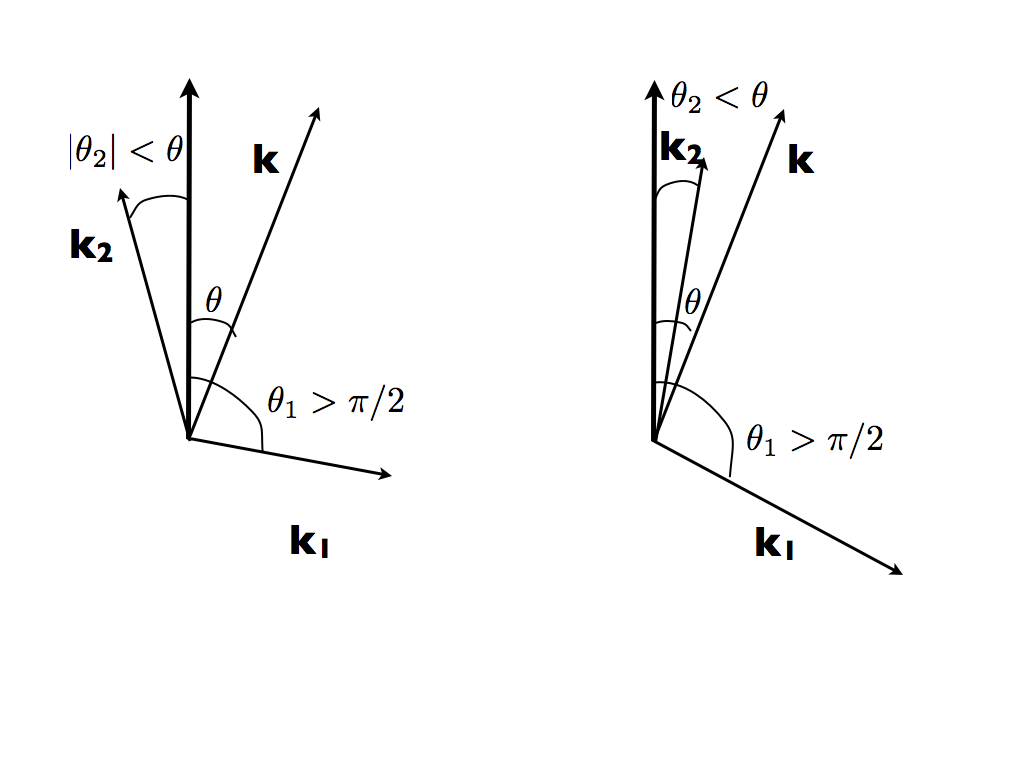}
\caption{ Examples of wave decay. Parallel propagating into two oblique modes with  $k_1=k_2 =k$. Obliquely propagating into forward modes, Obliquely propagating into one backward propagating modes. }
\label{pictures}
\end{figure}

 \subsection{Confluence of waves, $1;k,2$}
 \label{conf}
 
\begin{figure}[h!]
$
\begin{array}{cc} 
\includegraphics[width=.49\linewidth]{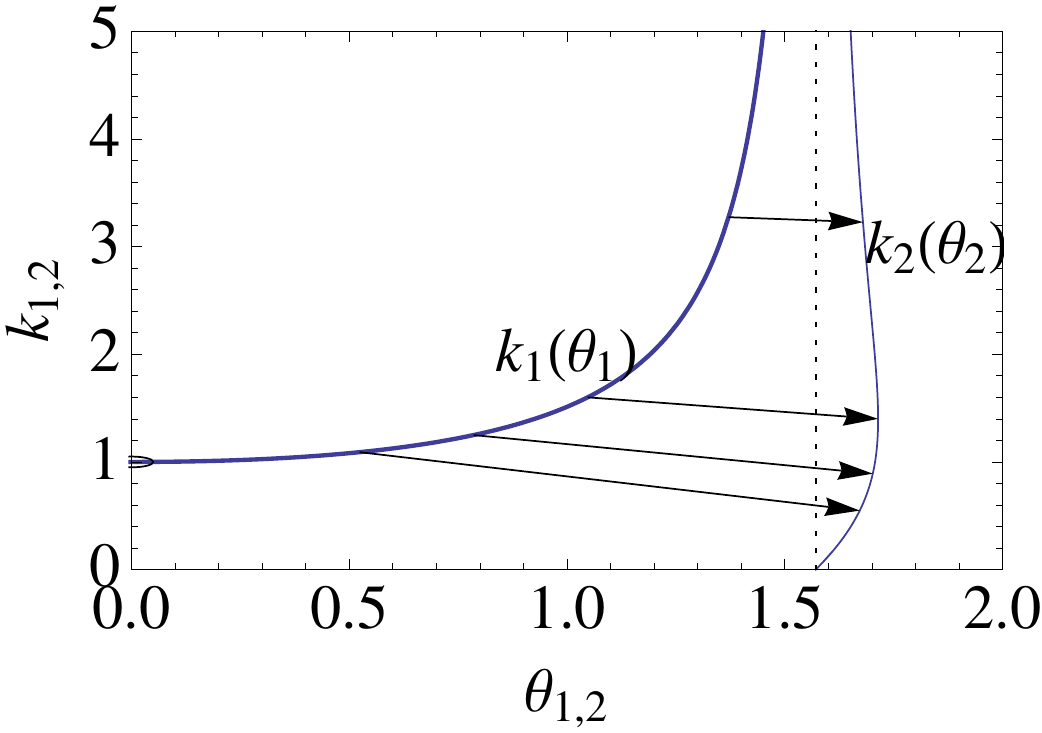} & \includegraphics[width=.49\linewidth]{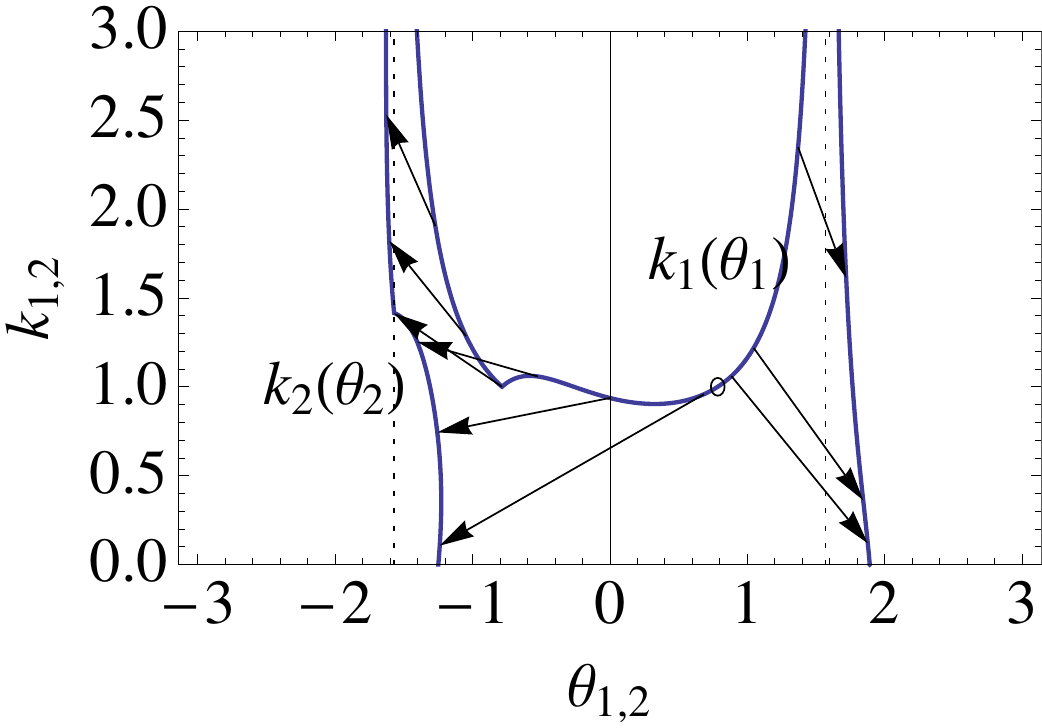}
\\
\includegraphics[width=.49\linewidth]{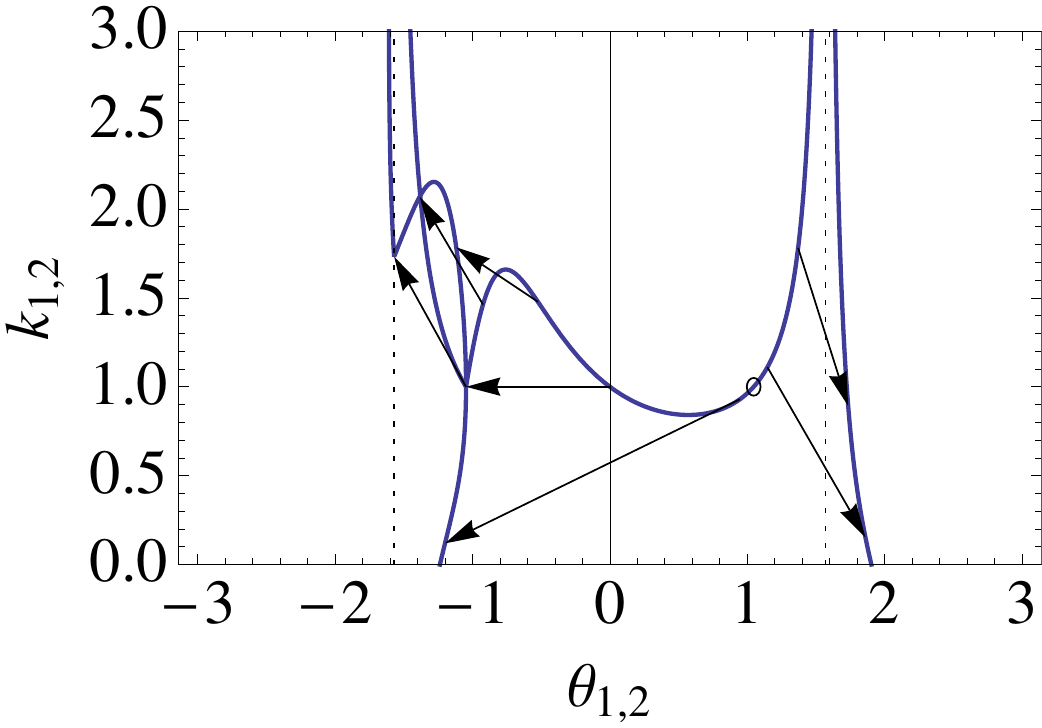}   & \includegraphics[width=.49\linewidth]{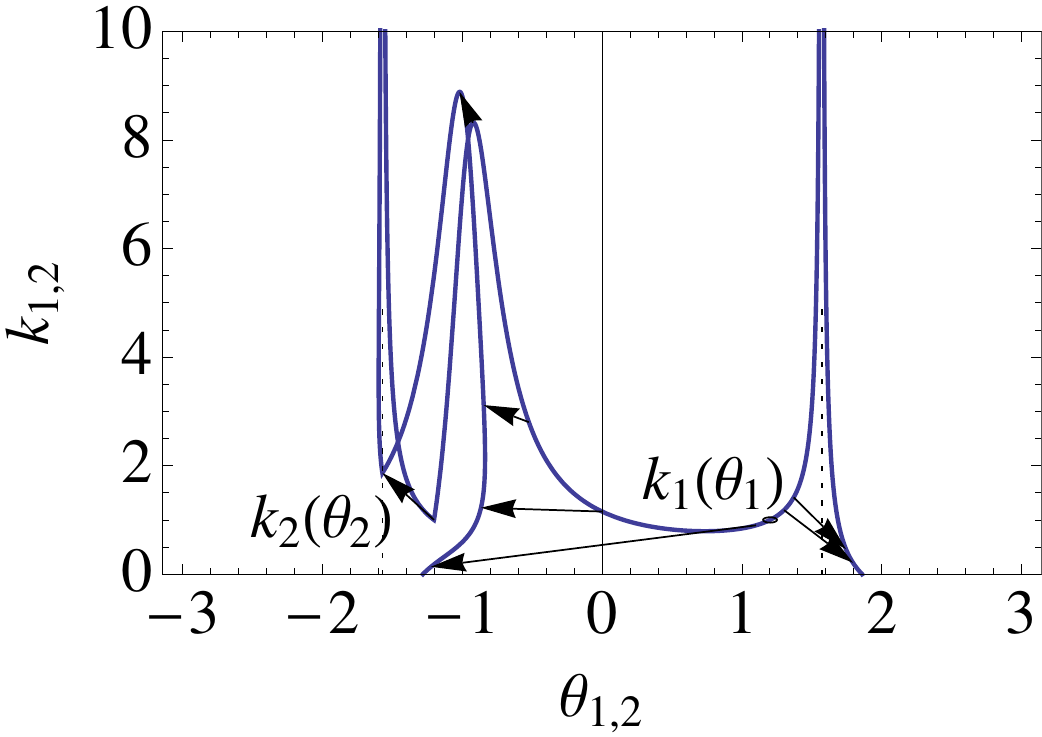} 
\end{array}
$
\caption{Resonant condition for $k;1,2$ interaction. {\it Upper Left Panel}: $\theta =0$. Both daughter modes can have arbitrary large $k_{1,2} \rightarrow \infty$ (for $\theta_{1,2} \rightarrow \pi/2_{\pm 0}$), one of the modes  propagates in the opposite direction, $\theta_2 > \pi/2$. For obliquely propagating parent wave $\theta > 0$ (other panels)  the resonant condition are very complicated even for co-planar propagation. For $-\theta < \theta_1 < \theta  $ the second mode has $ - \pi/2< \theta_2 < 0$ (when $\theta_1 \rightarrow -\theta_{+0}$, then $\theta_2  \rightarrow - \pi/2 _{+0}$). For $-\pi/2 <\theta_1 < - \theta$, then $\theta _2 < - \pi/2$.  The arrows connecting two daughter modes start at $\{k_1, \theta_1\}$ and end at  $\{k_2, \theta_2\}$. The two lower  panels show that the resonant condition for   $k;1,2$ interaction is highly sensitive to $\theta$: $\theta =0.66 \times \pi/2$ for lower left panel  and $\theta =0.765 \times \pi/2$ for lower right panel. }
\label{1k2}
\end{figure}

 In addition to the $k;1,2$ interaction, the kinetic equation (\ref{kinetic}) also involves the $1;k,2$ resonance. Now the two daughter modes $1,2$ are not interchangeable. The $1;k,2$ resonance is more complicated, Fig. \ref{1k2}.
 Fig. \ref{1k2} illustrates that the 3-waves processes couple modes with very different $k$ and $\theta$: 

The complications of the resonant conditions for $\theta \rightarrow \pi/2$ make even the kinetic calculations very  challenging: the resonant conditions change drastically on scales of $\sim $ hundredth of a radian. At the same time the resonant wave vectors vary, basically from zero to infinity. 
 To take correct account of these factors requires  high resolution in both $\theta$ and $k$.

\section{Non-linear interaction of whistlers, wave kinetic equation and Kolmogorov-like spectrum }
\label{inter}

\subsection{Weak interaction of whistlers}

Following the weak turbulence approach, we assume that the wave amplitude $b$ is a slowly  varying function of time, $b(\epsilon t)$, where $\epsilon \ll 1$ is the wave amplitude expansion parameter. Using  the EMHD equation as a Lie derivative of \Bf\ along the current, Eq. (\ref{Li}),  multiplying by $ {\bf e}_B^\ast$, and expanding in 
 $\epsilon$
\ba &&
  {\partial _t b_\k}= { \delta {\cal H}_{int} \over \delta b^\ast_\k}
\nn &&
{\cal H}_{int}  = \int d\k_1 d\k_2  \left( 
 {1\over 2}  V_{k\leftrightarrow 1,2} b^\ast  b_1 b_2 \delta(\k  - \k_1 -\k_2)    + V^\ast_{ 1\leftrightarrow k,2}b  b_1 b_2^\ast  \delta(\k_1 - \k  -\k_2) \right) 
   \nn &&
    V_{k\rightarrow  1,2} = \sqrt{ \om_1 \om_2 \over  \om} \left\{  {\bf e}_B^\ast \cdot \left[\left((\k_1 \times {\bf e}_{B_1}) \cdot \k_2 \right) {\bf e}_{B_2} - (\k_2 \cdot {\bf e}_{B_1}) (\k_2 \times {\bf e}_{B_2})   + (1 \leftrightarrow 2)  \right] \right\}
\ea
where 
\be
\delta \B_{n}  = { b_n  \sqrt{\om_n}}  {\bf e}_{B_n} e^{- i ( \om_n t - {\bf k}_n \cdot {\bf r})} , \, n=\{0,1,2\}
\ee
\citep[see][Eq.  (2.1.3)]{Zakharov}.

The interaction matrix $V$ takes an especially simple  form for  co-planar propagation in the $x-y$ plane. Setting $\phi_n=0$,
\ba&& 
V_{0\leftrightarrow 1,2}=  \sqrt{ \om_1 \om_2\over   \om_k}  {k_1( k_2-k_1)\over  2  \sqrt{2}}    \,  \cos ^2{\theta -  \theta_1 \over 2}  \sin (\theta_1 - \theta_2) + (1 \leftrightarrow 2)  =
\nn &&
 \sqrt{\om_1 \om_2\over \om_k}  {k_1- k_2\over  2  \sqrt{2}} \left( k_1(1+\cos (\theta-\theta_1) )+k_2( 1+\cos (\theta-\theta_2) )\right)   \sin (\theta_1 - \theta_2).
\label{V1}
\ea
(For co-planar decay the  angles $\theta$ can be taken to vary  between $-\pi $ and $\pi$).

The form of (\ref{V1}) clearly shows that decay into two co-linear modes (and merger of  two co-linear modes), $\theta_1 = \theta_2$, is prohibited.
  More generally, the interaction highly disfavors decay into nearly aligned modes, as well as interaction of nearly aligned modes.  The form of the interaction matrix $V$ has important implications for the development of a turbulent  cascade. Most importantly,  whistler waves with highly different angles can experience a three-wave interaction.

\begin{figure}[h!]
\includegraphics[width=.49\linewidth]{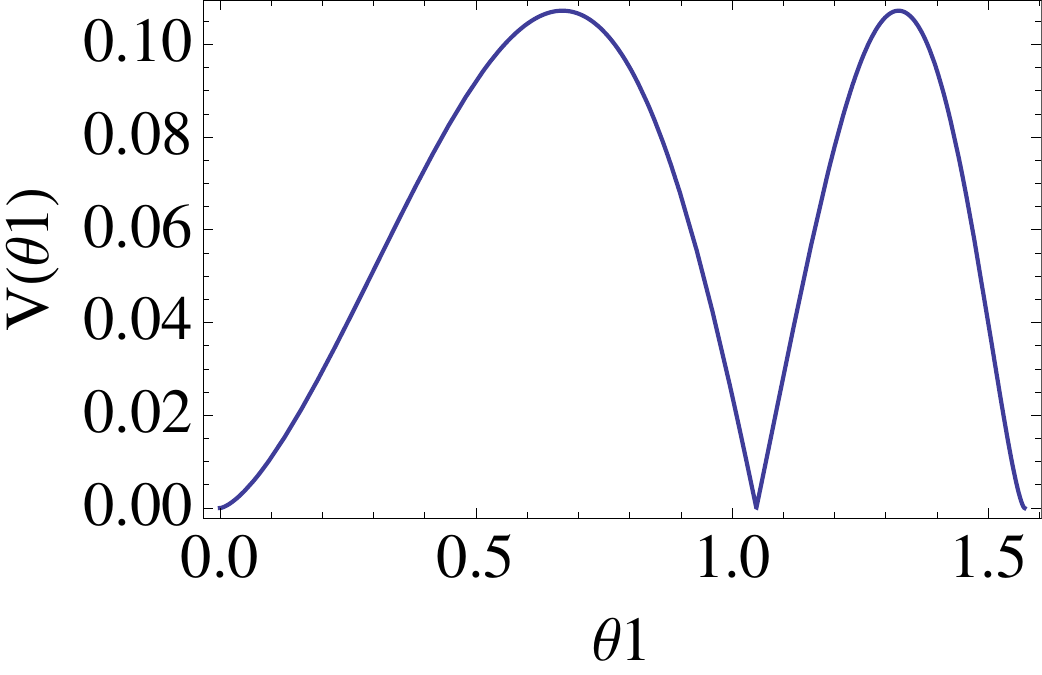}
\includegraphics[width=.49\linewidth]{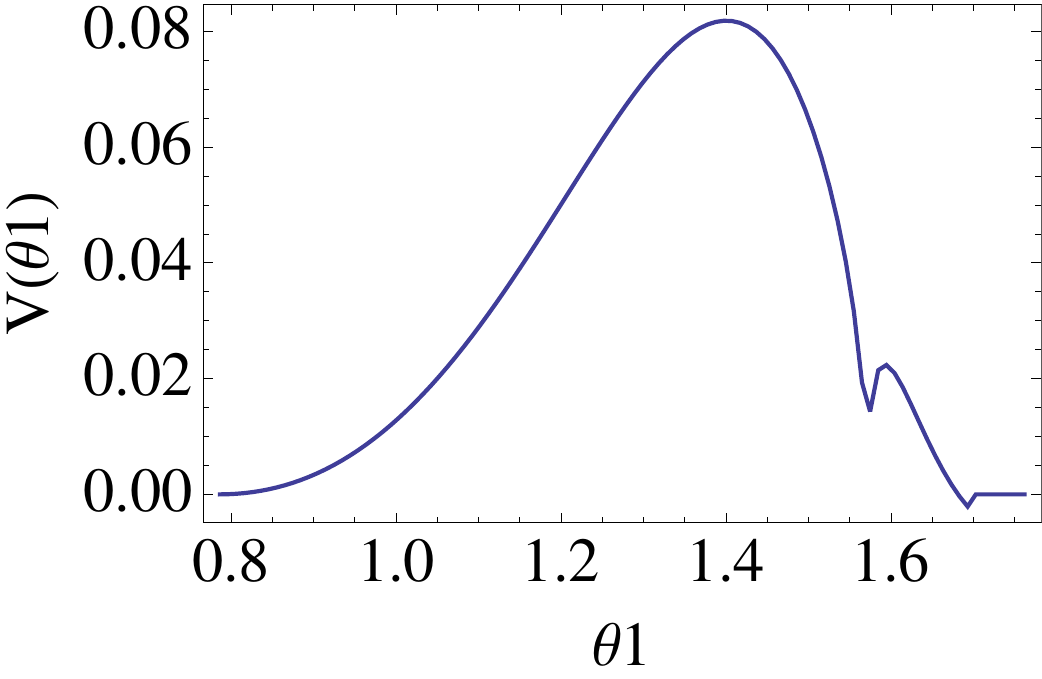}
\caption{{\it Left Panel}. Absolute value of the matrix element  $|V|$  for decay (\ref{V1}) for parallel propagating initial wave. Decay into two daughter modes with $\theta= \pi/3$ (and $k_1=k_2$ is prohibited).{\it Right Panel}. Same for the propagation of the initial wave at $\theta = \pi/4$. The dip is at $\theta_1= \pi/2$.}
\label{Voftheta}
\end{figure}

  For co-planar decay, for a  given initial wave $\{k, \theta\}$,  the interaction matrix depends only on one parameter (\eg\  the polar angle of one of the modes $\theta_1$) - the other parameters can be eliminated using resonant conditions; explicit rations for $k_{1,2} (\theta_{1,2})$ and $\theta_2 (\theta_1)$ were  derived in Section \ref{Co-pla}, see Fig. \ref{Voftheta}. 
 For the decay of the wave that is initially propagating along \Bf, $\theta =0$,   there is another property of the interaction matrix. As we discussed in Section \ref{non-linear}, waves with $k_1 = k_2$ and $\theta_1 = \theta_2$ do not interact, or, alternatively, a decay into two such modes is prohibited.

For the given three-wave interaction matrix (\ref{V1}) the kinetic equation reads
\citep[see][Eq.  (2.1.12)]{Zakharov}.
\ba &&
\partial_t n = \pi \int  \left[ |V_{0\leftrightarrow 1,2}|^2 f_ {k\leftrightarrow 1,2}  \delta(\k  - \k_1 -\k_2)    \delta(\om  - \om_1 -\om_2)  +
\right.
\nn &&
\left.
2 | V_{ 1\leftrightarrow k,2}| f _{ 1\leftrightarrow 0,2}  \delta(\k_1 - \k  -\k_2)  \delta(\om_1 - \om  -\om_2) \right]
d\k_1 d\k_2
\nn &&
 f_ {0\leftrightarrow 1,2}= n_1 n_2 -n  (n_1+n_2)
 \label{kinetic}
\ea

\section{Kolmogorov-like whistler cascade}

\subsection{Anisotropy}

Vainshtein probably  was the first person to consider  the EMHD turbulence  \cite{1973JETP...37...73V}. His reasoning relied on what is now called the  critically balanced cascade. This is  now  the most common  approach to treat the EMHD cascade  \citep{1999PhPl....6..751B,2003PhPl...10.3065G,2004ApJ...615L..41C,  2008PhRvL.100f5004H}. Following the development of the MHD turbulence by  \cite{1994ApJ...432..612S}, the MHD ideas of critical balance were transplanted to the EMHD turbulence. On the other hand, 
 \cite{Kingsep,RG} argued that the whistler turbulent cascade  is  isotropic and  weak.  
 
Above, we argued that the idea of the critically balanced cascade, based on the 4-mode interaction of non-dispersive \Alfven waves is not applicable to the EMHD cascade. 
On the other hand, a  number  of numerical simulations  \citep[\eg][]{2004ApJ...615L..41C,2009ApJ...701..236C,2012PhPl...19e5901T} seem to confirm the anisotropy, though the results of the numerical simulations regarding the spectrum remain controversial \citep{2012ApJ...758L..44B}. As we discuss below, observation of the highly anisotropic cascade  in numerical simulations is likely to be an artifact of the limited dynamical and temporal range of simulations.

 In Ref. \cite{2003PhPl...10.3065G}   analytic arguments  were suggested that the whistler cascade should be anisotropic. This was based on applying the  Zakharov-Kuznetsov transformation \citep{Zakharov}  to the kinetic equation.
Note, that 
generally, the matrix element for 3-wave  whistler interaction is not a separable function of $k_\parallel$ and $\k_\perp$, so that the Zakharov-Kuznetsov transformation cannot be achieved. In Ref. 
\cite{2003PhPl...10.3065G} it was  first assumed that the turbulence is highly anisotropic, nearly transverse; in this limit the matrix element becomes separable in $k_\parallel$ and $\k_\perp$, giving the scaling of the anisotropy cascade. In addition, the spectra suggested in Ref.  \cite{2003PhPl...10.3065G}  are not unique and  IR divergent \cite{boldyrev95}; this further raises questions about their applicability. As we discuss below, the nature of the interaction matrix is such that it  couples modes with very different propagation angles. As a results,  in equipartition the angular distribution could be quasi-isotropic.

In astrophysical  applications,  for example, the EMHD cascade can occur in a collisionless plasma as a continuation of the MHD cascade.  In Ref. \cite{1999ApJ...520..248Q} it was argued  argued that the turbulent energy which is not damped on scales larger than the proton Larmor is transformed into whistlers and is eventually damped on electrons. Thus ``stirring'' at the largest EMHD scale is likely to be highly anisotropic, continuing the highly anisotropic criticality balance MHD cascade. Then two somewhat separate questions in relation to the anisotropy of the cascade arise. First is the temporal evolution of the cascade that is being stirred, possible anisotropically, at the outer scale. Second is the asymptotic structure, when the cascade  reached the inner scale (along all directions for the anisotropic case). As we argue below these two case, time dependent and  time asymptotic structure of the cascade imply very different spacial structure.

\begin{figure}[h!]
\includegraphics[width=.99\linewidth]{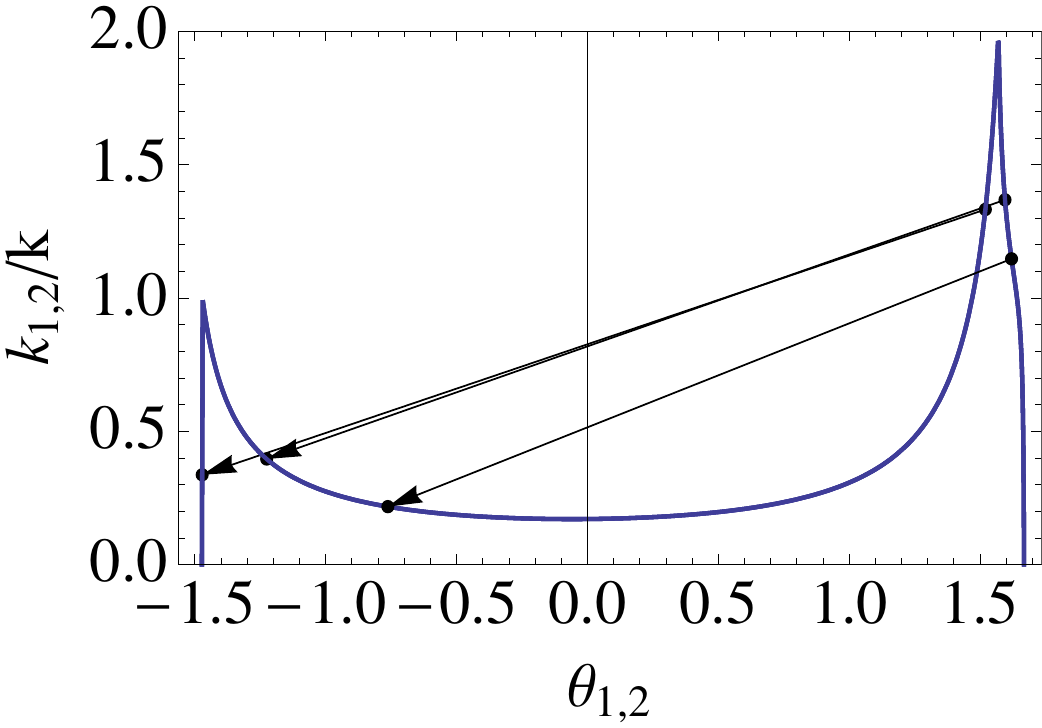}
\caption{Illustration that highly oblique whistlers decay preferentially into oblique whistlers, and that the interaction of two highly oblique whistlers produces a  highly oblique whistler, $\theta = \pi/2 -.1$. For the product with small angle, $\theta_{1,2} \rightarrow 0$, the corresponding wave number is small. }
\label{3wavwWhistlerhighlyoblique}
\end{figure}

First note, that  oblique mode decays  preferentially into  highly oblique, high $k$ mode, (see Fig. \ref{3wavwWhistlerhighlyoblique}).  The daughter mode with $k_{1,2} > k$ propagates at more oblique angles, $\theta_{1,2} > \theta$. Thus, each decay process will produce more oblique waves with higher $k$.
On the other hand,  the interaction of oblique whistler typically produce parallel modes with much smaller $k$.  Thus, if there were no inverse process of wave merger and confluence, {\it a simple wave decay will lead to highly oblique modes with}  $k \gg k_{inj}$. 

\begin{figure}[h!]
\includegraphics[width=.99\linewidth]{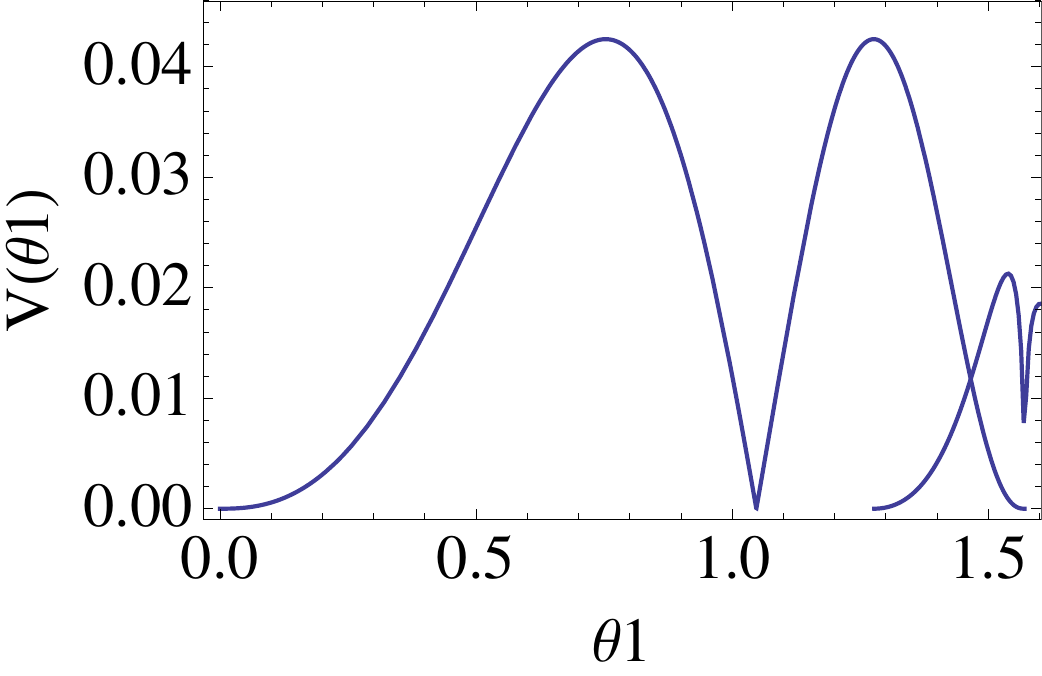}
\caption{Comparison of the decay matrix elements $|V|$  for decay (\ref{V1}) for parallel propagating initial wave, and of the daughter wave emitted at maximum of $V$, $\theta_1 = 1.27$. The most probable emission angle for the second generation decay product is within few percents of $\pi/2$.}
\label{1-2}
\end{figure}

Let us discuss the  wave decay in more details. 
The suppression of the matrix element for highly oblique daughter waves, $\propto \sqrt{\om_{1,2}} \propto \sqrt{\cos \theta_{1,2}}$, is partially compensated by large phase volume, $\propto  \theta_{1,2}$. A single wave initially propagating along the \Bf\ becomes nearly orthogonal within two decay cycles. For example, a parallel propagating mode has a decay probability peaking at 
 $\theta_{1,2} = 1.27$. The secondary mode, propagating at $\theta = 1.27$ has a decay probability peaking at $\theta_{1,2} =1.53$, see Fig. \ref{1-2}. Thus, in the time-dependent case, if energy is injected at the outer scale of EMHD cascade, it will quickly within two interaction times become quasi-two-dimensional   at $k\gg k_{inj} $  just due to simple non-linear decay of  whistler modes. By the nature of limited time evolution in numerical simulations, this is, we suggest, the origin of the anisotropy  observed by a  number of collaborations. It's a reflection of the fact that a simple non-linear whistler decay proceeds mostly in the perpendicular direction. 
 On the other hand, at wave length not much larger than the injection, there is an inverse process of wave merging, that generally leads to the isotropization of the turbulence.

Note, that {\it the whistler turbulence cannot become purely two dimensional:} for $k_\parallel = 0$ the non-linear interaction freezes out: {\it  any configuration $B_z (x,y)$ satisfied the stationarity condition.} Thus, a cascade must develop in both directions, along the field and perpendicular. This also implies that the works which considered purely two dimensional EMHD turbulence need to be reconsidered.

 The  considerations given above are generally confirmed by numerical simulations, Section \ref{Solving}, though a steep frequency dependence on the wave number, quickly increasing cascade rate and the apparent subtle dependence on the initial conditions prevent a decisive conclusion.

Finally, we comment on the applicability of the reduced MHD approach  to EMHD.
For strong axial fields there  
is an often used approach called reduced MHD
 \cite{1992JPlPh..48...85Z,1997noma.book.....B}, which allows considerable simplifications to the dynamic equations. Under reduced MHD approximation, the guiding \Bf\ is assumed to be constant, while  the  fluctuations of the \Bf\ are typically parametrized via the flux function  $P$: $\delta \B_\perp = \nabla (P) \times {\bf e}_x$. This approach has also been applied to EMHD and  closely related kinetic Alfv{\'e}n wave turbulence, \eg\ Ref. 
 \cite{2012PhPl...19e5901T,2012ApJ...758L..44B}. 
 
 The reduced MHD approach is not applicable to EMHD.
 Whistlers are inherently three-dimensional structures: an obliquely propagating whistler mode produces field fluctuations in all directions, which are generally comparable in amplitude, 
 $\delta B_\parallel \sim \delta B_\perp$. In fact, {\it  a highly oblique mode produces large  $\delta B_\parallel$}, see Eq. (\ref{whistlers}).    It is also  easy to see that using the reduced MHD parametrization mentioned above  the EMHD equations (\ref{main}) cannot be satisfied.

\section{Numerical solution of  the kinetic equation}
\label{Solving}

In this Section we numerically  integrate Eq (\ref{kinetic}). We assume that at small wave numbers a large number of whistler modes is injected.  We  assume that the turbulence is back-forth symmetric, so that we can limit our integration to $0< \theta <\pi/2$ (if a wave with $\theta > \pi/2$ is generated, it is then assumed that the counter propagating cascade generates a similar wave in the  $0< \theta <\pi/2$ range).

Integration is done according to the following algorithm. For co-planar propagation three of the four integrations  (over  $\{k_1, \theta_1\}$ and  $\{k_2, \theta_2\}$) are resolved using the resonance conditions (\ref{kjk}-\ref{kjk1}). Special care should be taken to account for the allowed combinations of the resonant modes and discarding the non-physical solutions, as described above. 
This leaves only integration over $\sin \theta_1 d \theta_1$. 
 For a given $\{k, \theta\}$ we sum  is over all   allowed $\theta_1$ both for $k;12$ and $1;k2$ processes. The time step is adjusted so that the relative variations of the occupation numbers at largest $k$ be smaller than unity.  Modes that fall into range $k>k_{max}$ are removed. In addition, modes in the last $\theta$-cell are removed, as discussed above any $B_x(y,z)$ is a solution of the stationary EMHD equation  and thus do not participate in the cascade.

\subsection{Code validation}

First note, that as we discussed in Section \ref{Co-pla}, highly oblique low-$k$ modes can quickly create very high $k$ modes. Keeping track of these  high-$k$ modes is difficult, both due to their high wave-number (this puts constraint on the simulation domain) and due to the fact the interaction strength, $ \propto V^2$ grows as $ \sim k^6$. As a result,  at high $k$ the cascade becomes very fast,  so that  at the upper limit on wave numbers  the  change in the occupation numbers becomes of the oder of unity in one time step, possibly violating the assumed weak interaction, while it is still slow at low $k$. This is also related to the fact at sufficiently high $k$ the cascade becomes explosive (see below). This puts a constraints on the integration time step.  This, combined with the fact that even in the first generation the non-linear interaction produces very high $k$ modes, does not allow to search numerically for a steady-state distribution. 

\begin{figure}[h!]
\includegraphics[width=.49\linewidth]{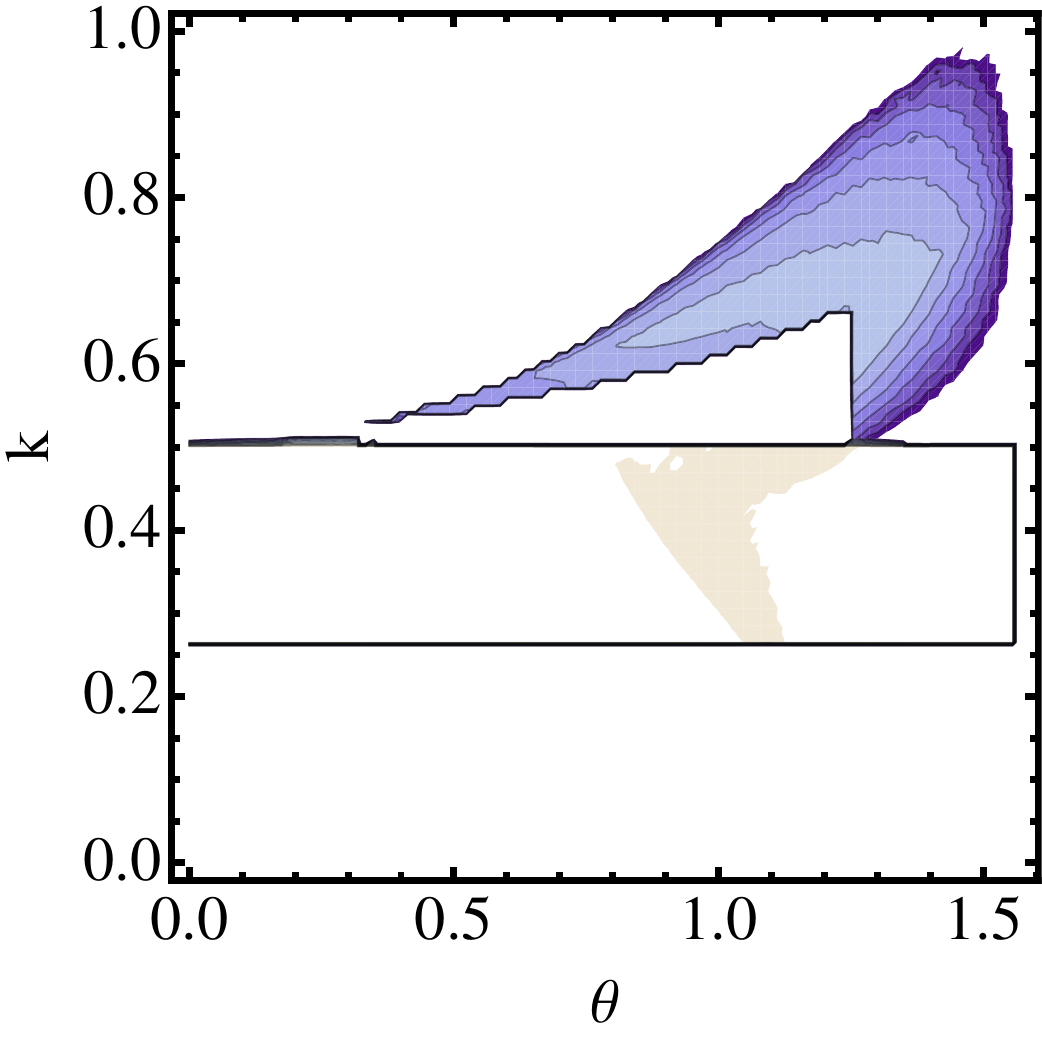} 
\caption{Detailed simulation of the first generation spectrum close to the injection scale. Note that, first, at $k \geq k_{inj}$ the spectrum is highly anisotropic.  Often, this is  a transient effect, see text for details. }
\label{EMHD-Turb}
\end{figure}

\subsection{Results of the numerical solution of the kinetic equation}

In this Section we discuss the results of the numerical solution of the kinetic equation. In the end
we conclude that numerical integration confirm our analytical expectation: transiently, the wave decay leads to the  formation of  a highly oblique distribution at $k > k_{inj}$. Later, when a broad spectrum in $k$ is present, the distribution can become  a quasi-isotropic. 

Generally, the simulations are highly dependent on the initial
conditions. Depending on spectral properties of the injection (width in 
$k$ and  $\theta$), behavior of the cascade can be drastically different. 
  Generally broader injection results in faster cascade, while narrow 
injection may, under certain conditions, produce no cascade (within 
the integration limit). 
  Injection spectra differing by a factor of a few in angular width 
could have dynamic time scales different by orders of magnitude. 
Injection spectra limited in $\theta$ have strongly suppressed
evolution and may  not develop  a cascade. 

\begin{figure}
\includegraphics[width=.99\linewidth]{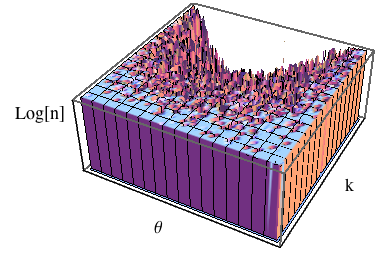} 
\caption{Example of the non-linear evolution after one wave interaction. Initially, a constant number of waves per unit range of $\theta$ and $k$ is injected in the whole integration domain. The figure shows that higher $k$ modes evolve much faster than the small $k$ modes. This puts constraints on time integration and the overall range of the simulation. Due to complicated resonant conditions the wave interaction is intermittent, especially at small $k$. For injection homogeneous in $k$ and $\theta$ the modes are preferentially scattered out of small angles of propagation.}
\label{intermit}
\end{figure}

Due to complicated resonance conditions the spectra are highly intermittent in $k$ and $\theta$; the  intermittency depends on the resolution. This is especially true at the edges of the integration domain, Fig. \ref{intermit}. 

Let us  discuss a number of calculations that concentrate on the first generations of the cascade development.
 Due to sensitive dependence of the interaction on the spectrum the time steps of the plotted examples are not equal - they are adjusted to display a typical interaction time scale.

Simulations of the broad angular injection spectrum are especially challenging: for sufficiently broad range of wave vectors even at first time step waves with the injection domain can interact leading to very high transfer rate, and putting very tight constraints on the integration time step. 

First, we inject whistler modes in  a limited range of wave number close to the lower limit of the calculation domain and consider various angular injection spectra:   isotropic, $\theta_{inj} < \pi/4$ and  $\theta_{inj} > 3\pi/4$.

\begin{figure}[h!]
$
\begin{array}{cc}
\includegraphics[width=.5\linewidth]{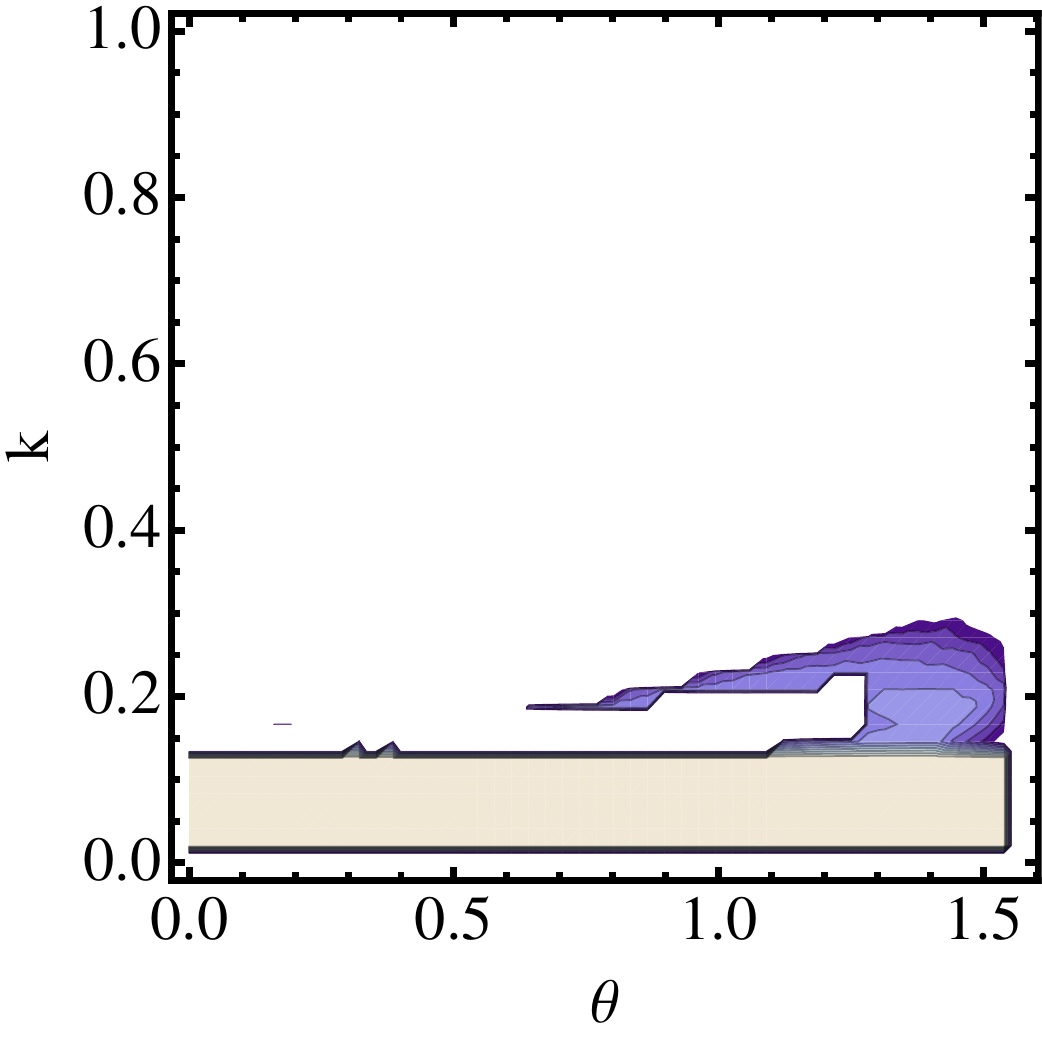} & 
\includegraphics[width=.5\linewidth]{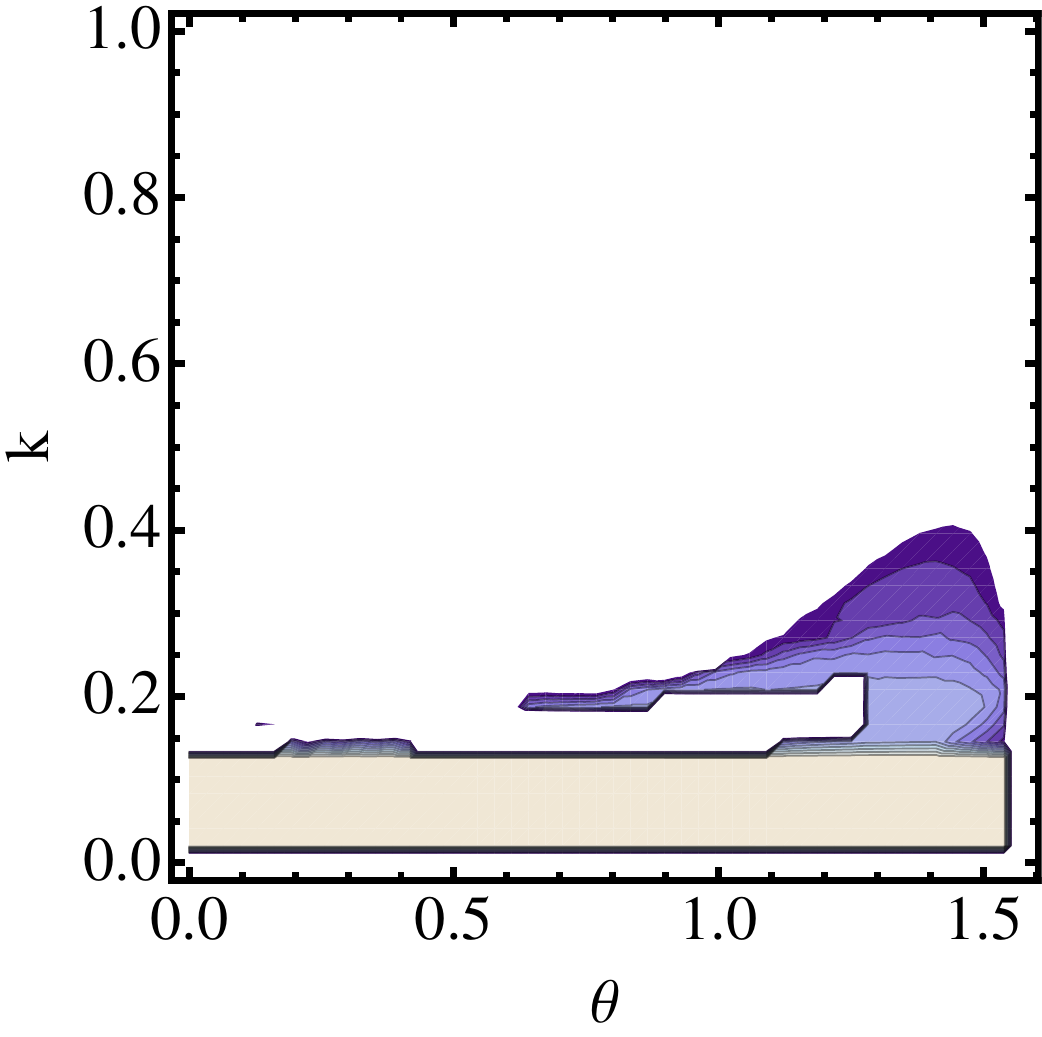}  \\
 \includegraphics[width=.5\linewidth]{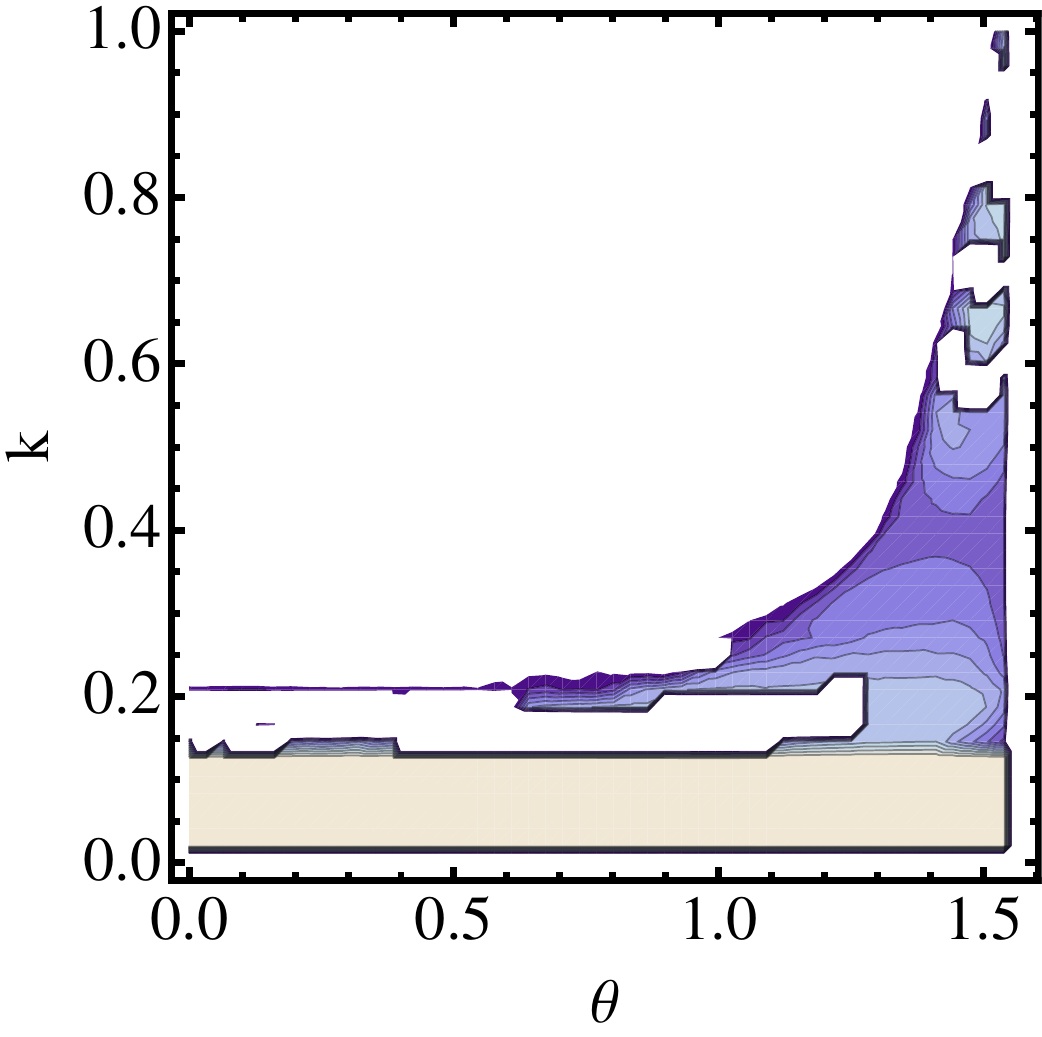}  & \includegraphics[width=.5\linewidth]{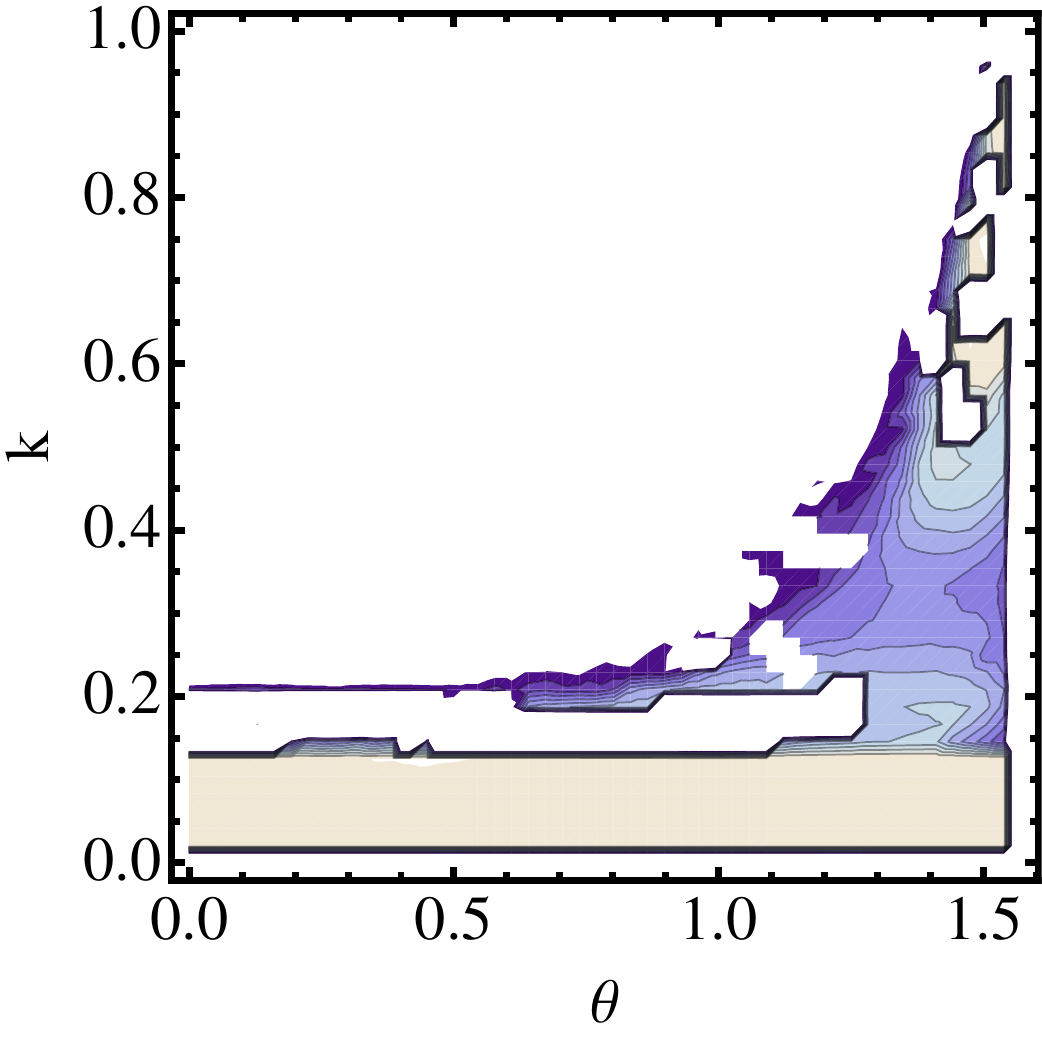} \end{array}
$
\caption{Typical development of an initially isotropic cascade. Waves are injected  with constant rate at small $k_{inj}$, at $0< \theta<\pi/2$ and are damped at $k> 10$. The spectrum  reaches a steady state which is  highly anisotropic. }
\label{iso}
\end{figure}

\begin{figure}[h!]
$
\begin{array}{cc}
\includegraphics[width=.5\linewidth]{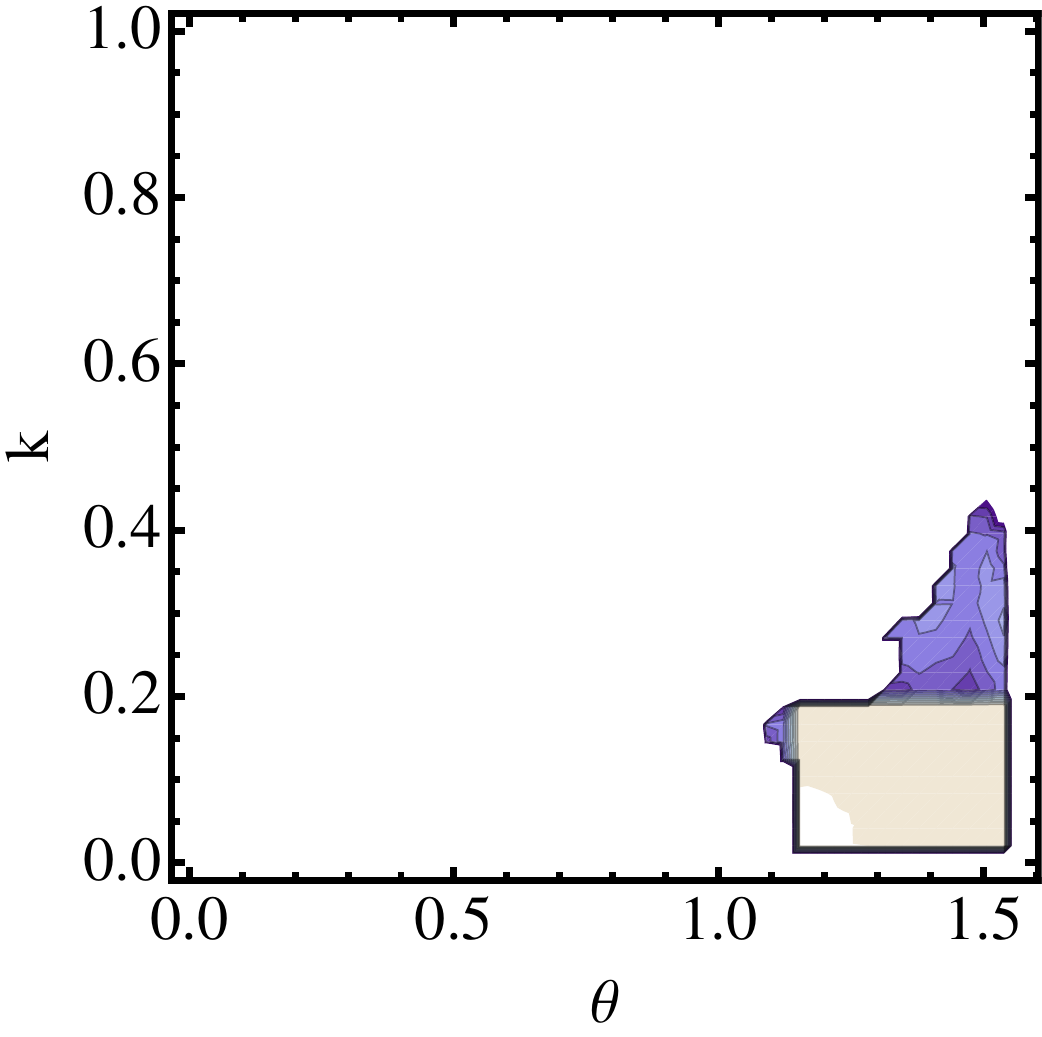} & 
\includegraphics[width=.5\linewidth]{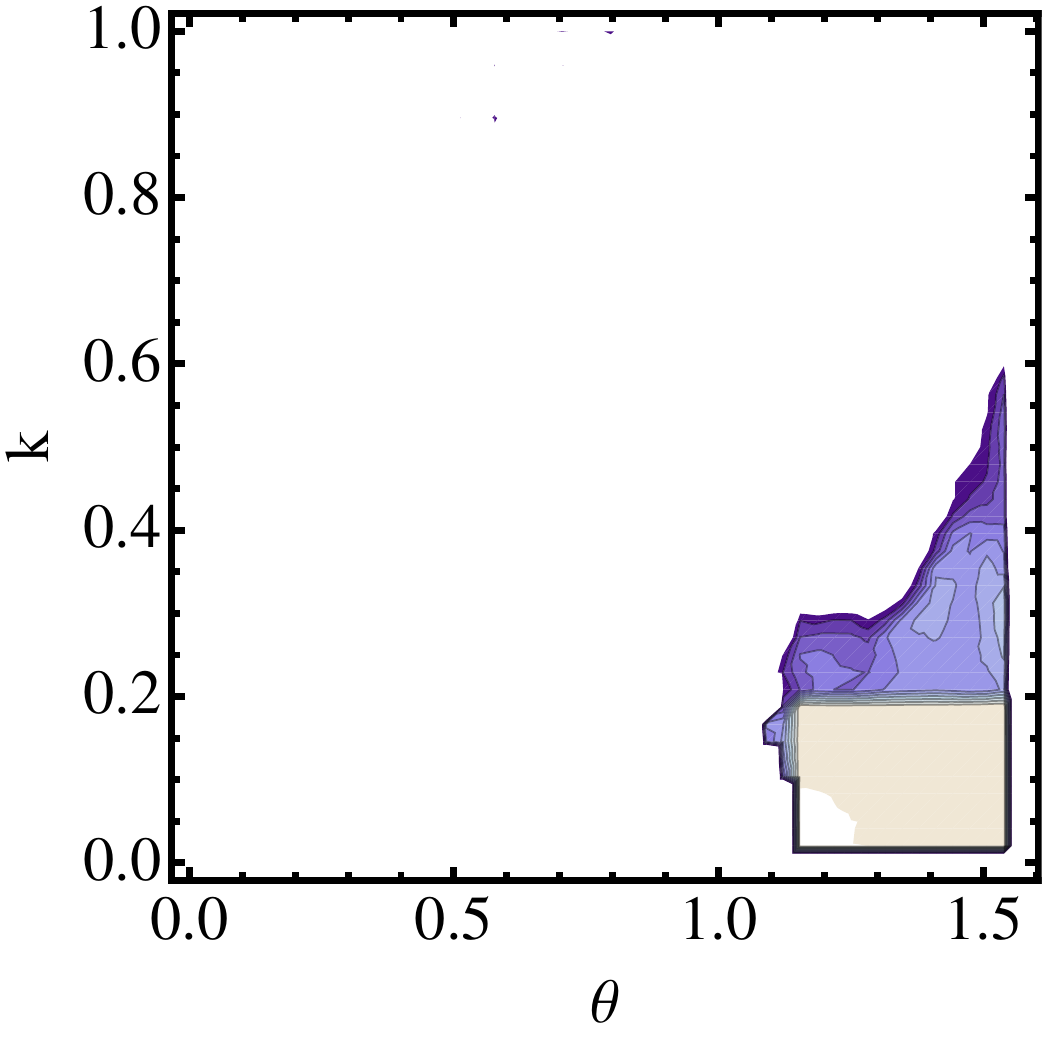}    \\
  \includegraphics[width=.5\linewidth]{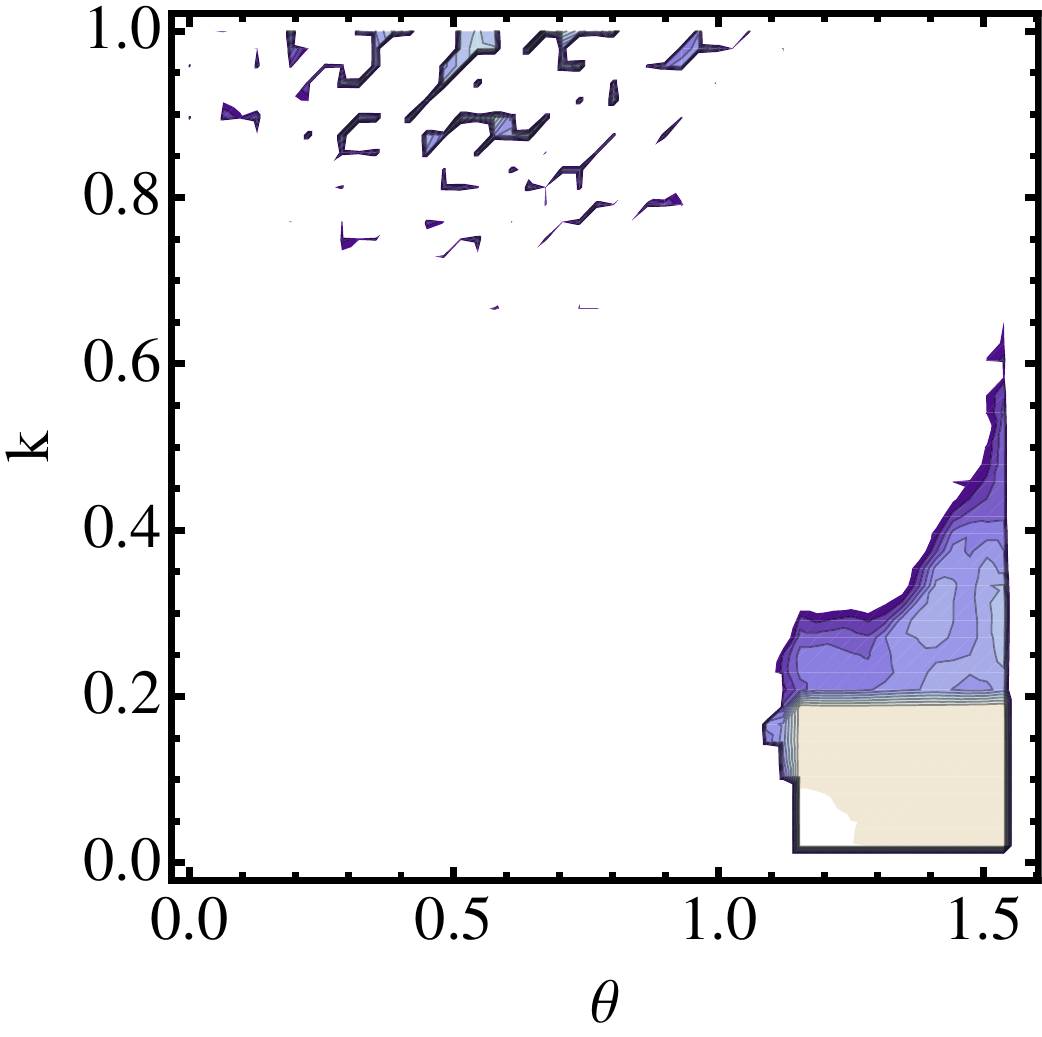}  &
    \includegraphics[width=.5\linewidth]{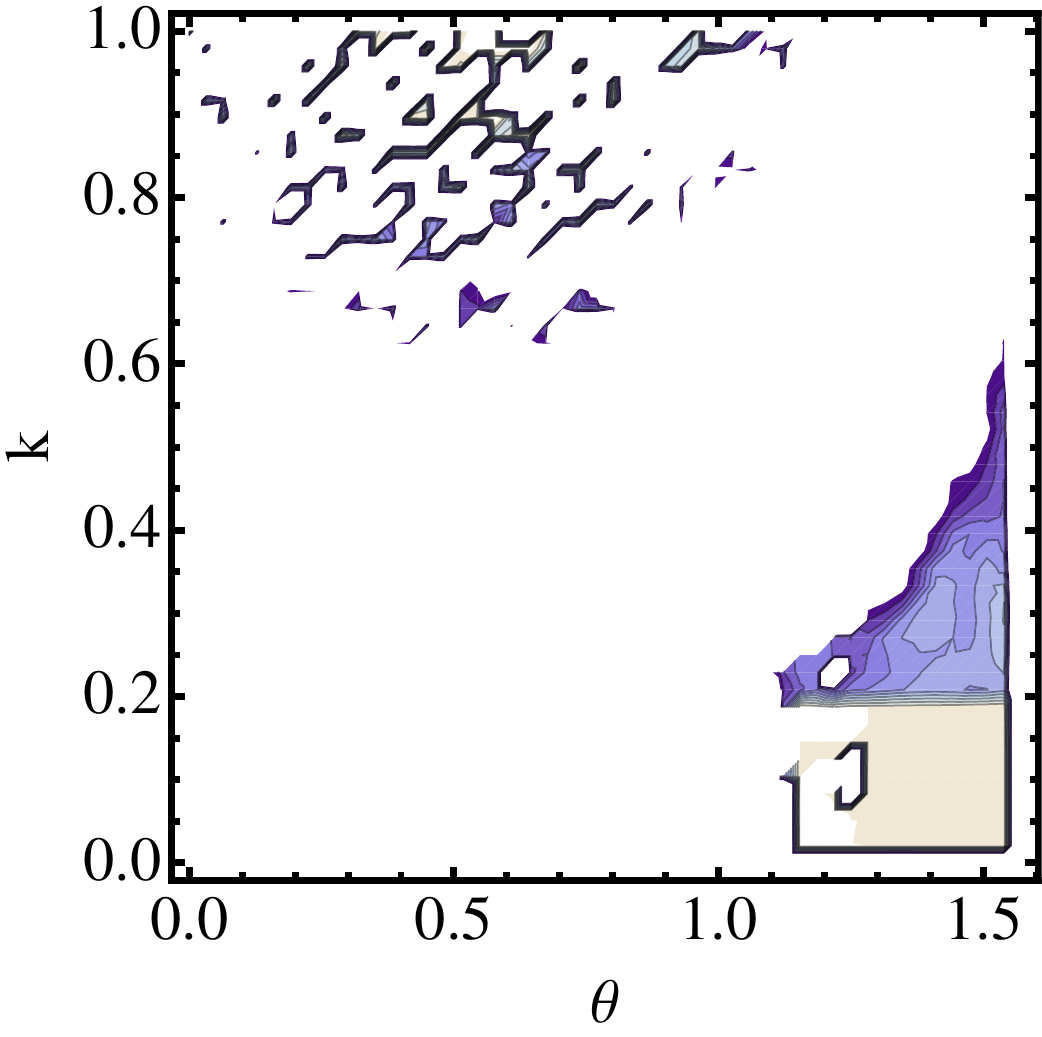} 
      \end{array}
$
\caption{Typical development of an initially quasi-perpendicular  cascade. Waves are injected  with constant rate at small $k_{inj}$, at $\pi/4< \theta<\pi/2$ and are damped at $k> 100$.  The distribution at high-$k$ becomes parallel-enhanced. }
\label{Large-theta}
\end{figure}

\begin{figure}[h!]
$
\begin{array}{cc}
\includegraphics[width=.5\linewidth]{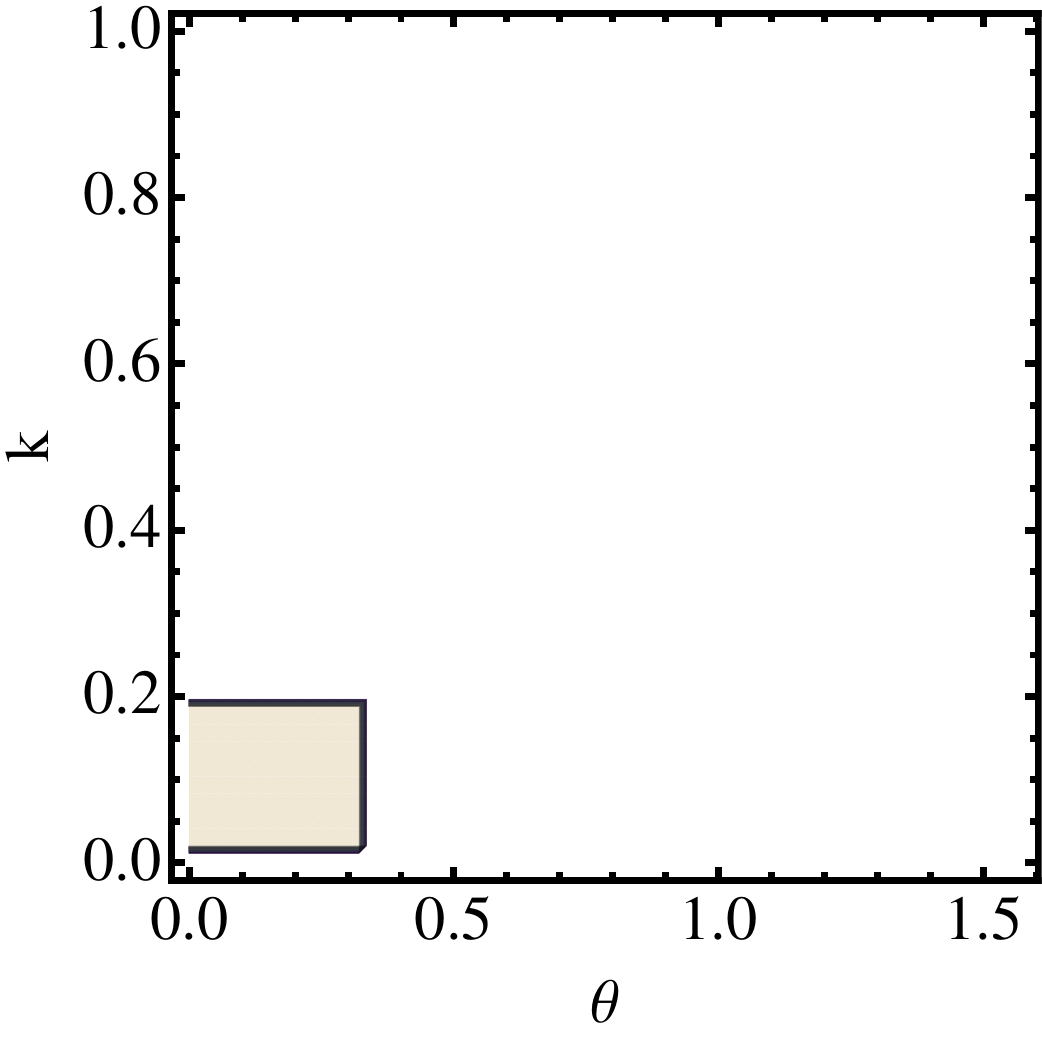} & 
\includegraphics[width=.5\linewidth]{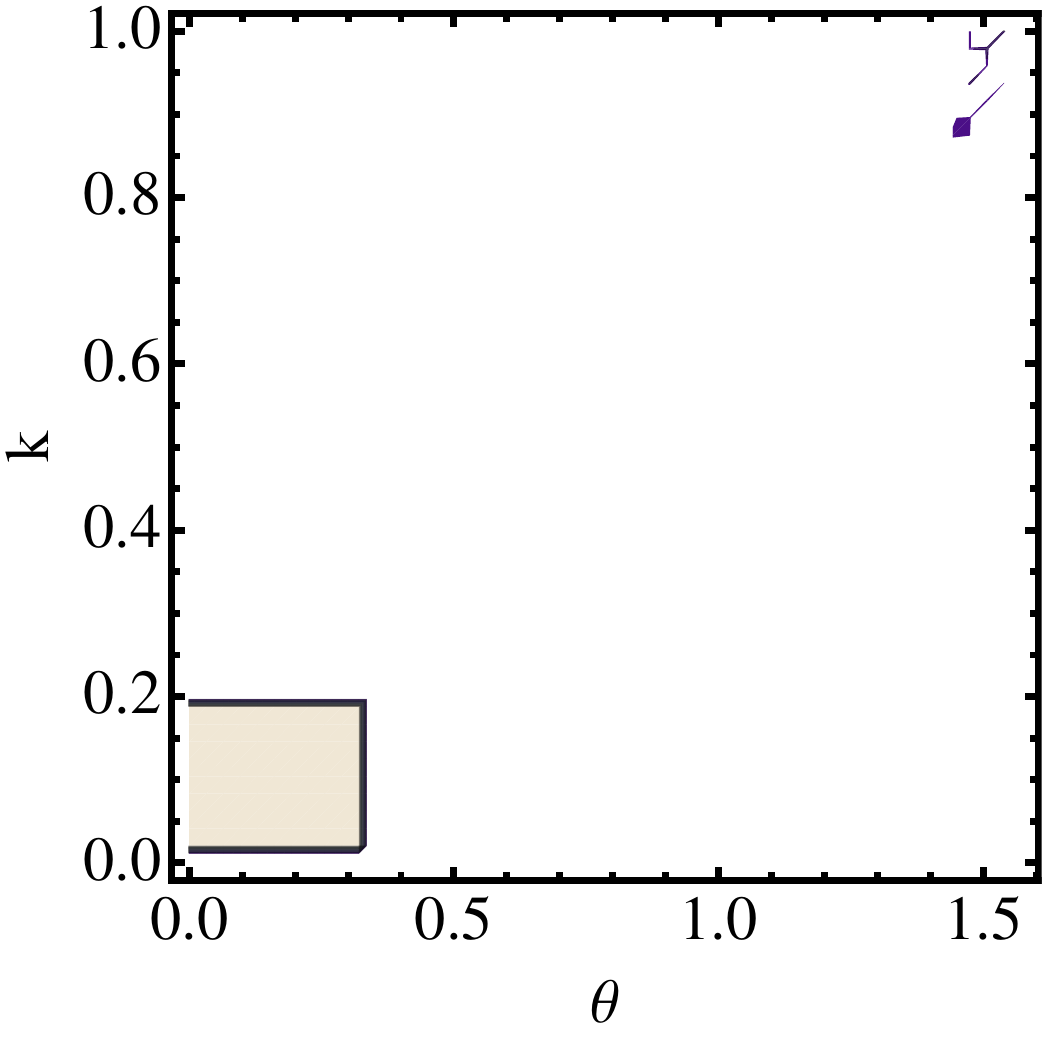}  \\
 \includegraphics[width=.5\linewidth]{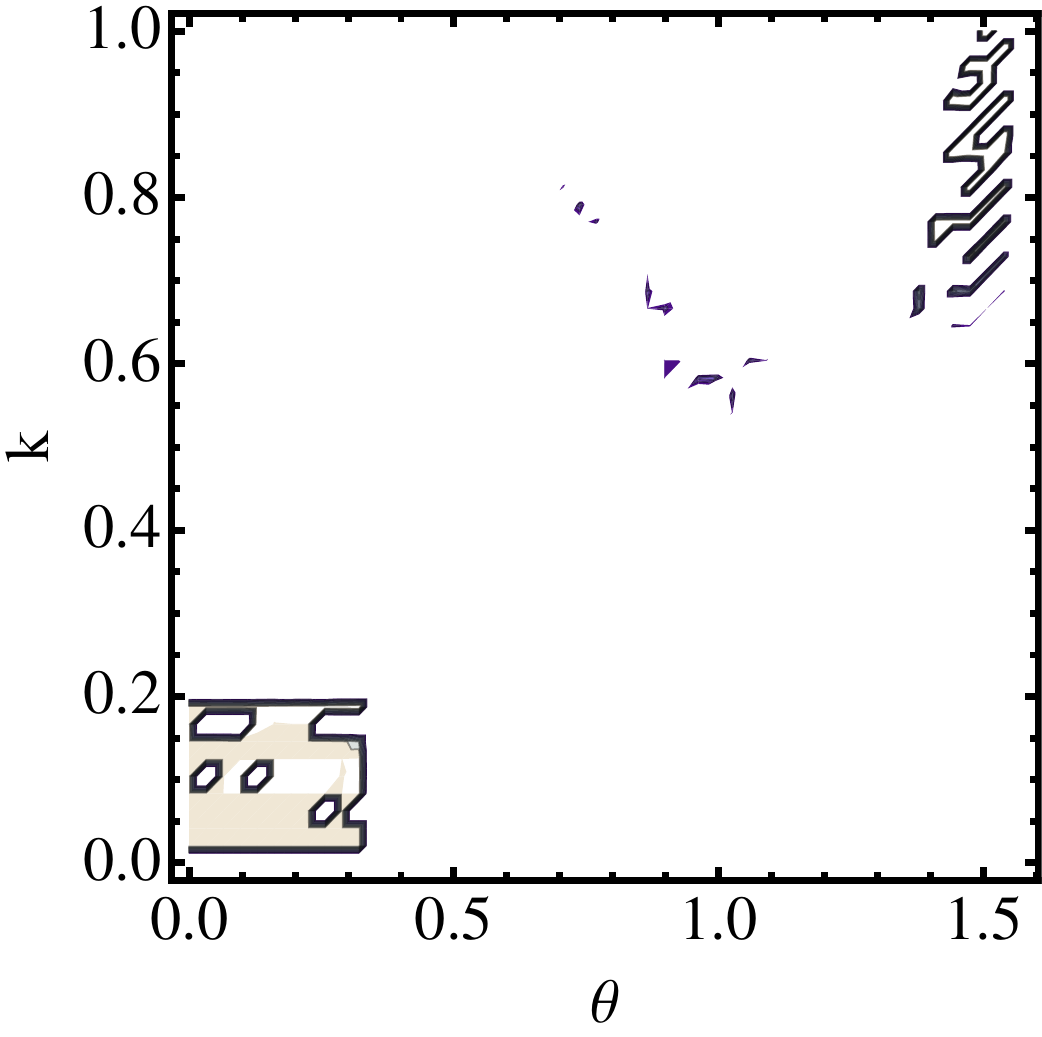}  & \includegraphics[width=.5\linewidth]{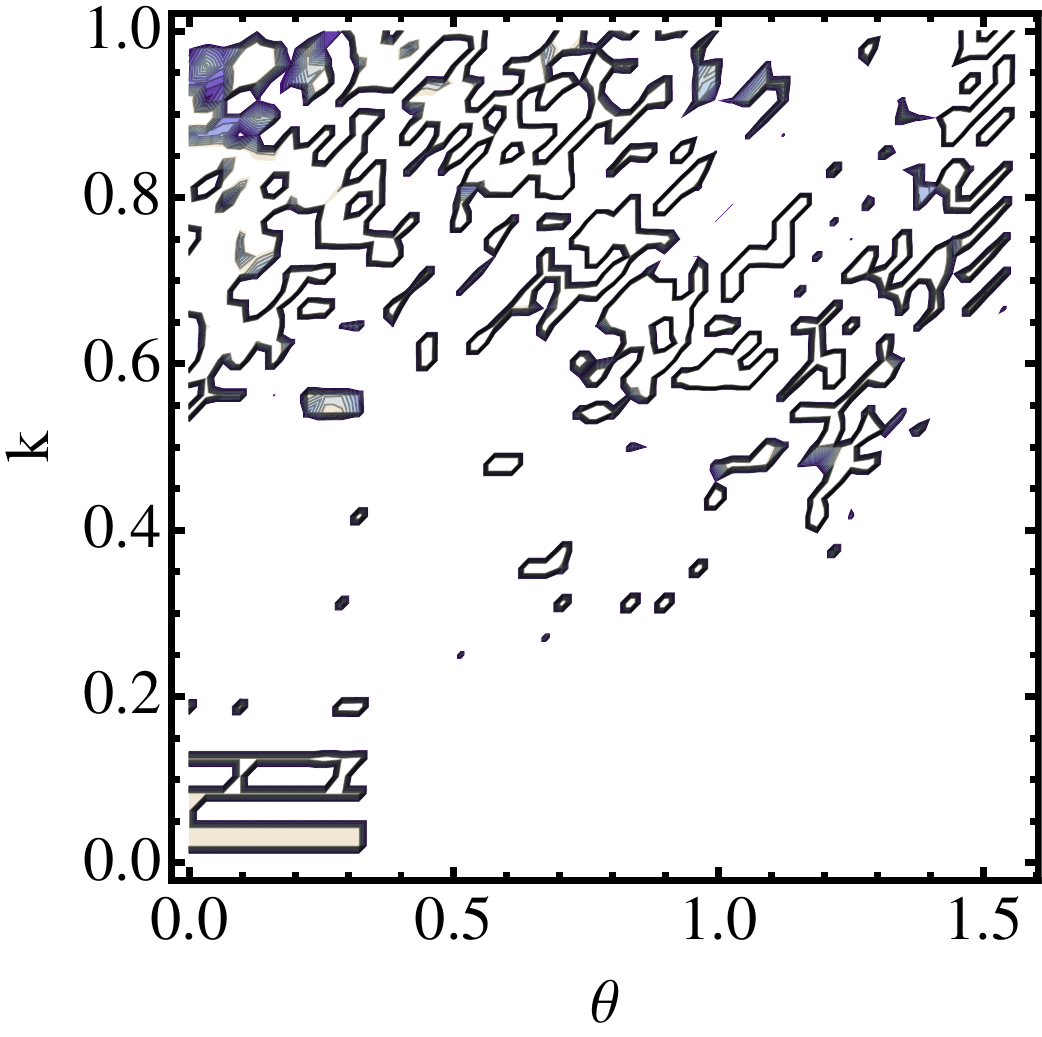} \end{array}
$
\caption{Typical development of an initially quasi-parallel cascade. Waves are injected  with constant rate at small $k_{inj}$, at $0< \theta<\pi/8$ and are damped at $k> 10$.
Initially the cascade develops very slowly (integration time scale here is orders of magnitude longer than in Fig. \ref{iso}), building a population of high-$\theta$ mode; after that the development is similar to the quasi-perpendicular injection: high-$\theta$ mode create a broad population of medium-$\theta$ modes. Plotted time steps are number 9,10,11,12 -  at shorter times there is appreciable evolution while at the last time step  the evolution become too quick to be captured by kinetic  simulations.
 }
\label{small-theta-1}
\end{figure}

As expected from the simple three-wave decay, Section \ref{three-wave}, initially, at $k> k_{inj}$ the spectrum is dominated by the decay products of the injected waves. Those decay products propagate preferentially at large angles to the \Bf\ and may produce an impression of a highly anisotropic cascade.
 On the other hand, the second generation is nearly isotropic, presumably due to merger of two oblique waves of larger and/or similar $k$. 

We  also find that the development of the cascade is highly dependent on the details of the injected spectrum. For a given range of injected wave numbers $k_{inj}$, a broader injection spectrum in $\theta$ would produce in the first generation  more obliquely propagating waves.  A spectrum limited in $k$ and propagating at  small $\theta$ creates high-$\theta$  modes and vise versa, a spectrum limited to large $k$ create a broad, yet preferably parallel distribution, Fig. (\ref{Large-theta}). 

Based on a number of simulations runs we conclude that (i) given enough time the EMHD turbulence develops within a broad range of angles, {\it without} strong preference for perpendicular propagation; (ii) temporarily, especially for wave numbers not much larger than $k_{inj}$ and for particular angular injection spectra, the cascade  generally {\it does} develop high powers at large angles - but  this is a transient effect;  (iii) the development of a cascade is sensitive to the injection spectrum: for parallel-dominated injection the cascade {\it initially} develops slowly, yet given enough time  a substantial perpendicular component is generated; (iv) the cascade efficiently couples high and small propagation angles, and creates in one integration time step very low and very high wave numbers; (v) it is not clear if the cascade is stable:  often the development of the cascade becomes explosive, quickly generating very large occupation numbers (and, thus violating the approximations used to derive the kinetic equation).

\section{Kolmogorov-like spectrum of whistlers}
\label{Kolmogorov-like}

\subsection{Spectrum}
Above,  in Section \ref{Solving}, we argued that based on the numerical simulations  EMHD is  non-universal, the evolution of the cascade depends on the details of the forcing. Often,
given enough time the EMHD turbulence develops within  a broad range of angles, without strong preference for perpendicular propagation. The cascade cannot become truly isotropic, though. 

Let us  next estimate the parameters of the quasi-isotropic  EMHD turbulence. Changing to occupation number presentations, on dimensional grounds, 
\be
\partial_t b \propto {c \over n e} \sqrt{\om} k^2 b^2= V b^2
\ee
where $V$ is the  3-wave interaction coefficient.
Thus, the interaction matrix  $V$ scales as 
\be
V \approx {c \over n e} k^2 \sqrt{\om}  \propto k^3
\ee
The interaction matrix  $V$ has the  dimension $cm^{1/2}   g^{-1/2}  s^{-1/2}$, as it should \citep{Zakharov}.

Assuming locality in $k$-space and  isotropy, the conservation of energy flux requires $n_k \om_k \nu_k =  $constant. The non-linear transfer rate \citep[Eq. 2.1.13 of Ref. ][]{Zakharov} is 
$
 \nu_k  \approx V^2 n_k  /\om_k 
 $ where 
  $n_k$ is 1-D number density of the excitations.
  Using scalings of the matrix element,
 $
 V \propto k^m, \, m=3
 $, this gives
 \ba && 
 n_k \propto k^{ -m}  \propto k^{-3}
 \nn &&
 \epsilon_k= B_k^2  = \om_k  n_k \propto k^{-1}
 \nn &&
\delta B_k  \propto k^{-1/2}
\nn &&
E_k  \propto  \delta B_k ^2 /k \propto k^{-2}
 \ea
 where $E_k$ is the spectral energy density of magnetic fluctuations.

Let us introduce Hall times  on the outer scale  $L$
\be
t_H = { L^2 \om_p^2 \over c^2 \om_B}
\ee
The non-liner transfer frequency  is then 
\be
\nu_k t_H \sim ( k L)^2 {\delta B_k^2 \over B_0^2}
\ee
where $\delta B_k$ is a fluctuation at the scale $L$. This
agrees with  Ref.   \cite{RG}.

For this cascade scaling, the fluctuations of the \Bf\ are 
\be
\delta B_k = {\delta B_L \over \sqrt{ kL}},
\label{deltaB}
\ee
while the 
 non-linear transfer frequency  is 
 \be
 \nu_k t_H = k L {\delta B_L^2 \over B_0^2}
 \label{nuk}
 \ee

If at the outer scales  $\delta B_ L \sim  B_0$, 
then $t_H \nu_k \sim k L \gg 1$  for $ k  \gg 1/ L$.
So, the largest eddy lives Hall time at largest scale, smaller eddies 
live shorter.

The cascade will be damped resistively on small scales. Taking resistivity into account the EMHD equation (\ref{main}) becomes
\be
{\partial\B\over\partial t} = - {c \over 4 \pi  e} \nabla \times    \left({\nabla  \times \B  \over n} \times \B \right) + \eta \Delta \B
\label{main1}
\ee
where $\eta$ is magnetic diffusivity.
Let us introduce the Hall-Lundquist number $S_H$  as the ratio of the linear dynamic to dissipative terms in Eq. (\ref{main1})
\be
S_H = {\om \over \eta k^2} = { c^2 \om_B \over \eta \om_p^2} = {L^2 \over \eta t_H}
\ee
In Ref. \cite{RG} this parameter was called ${\cal R}_B$ and it was  argued that for $S_H \gg 1$ the Hall plasma evolves though a turbulent cascade with inertial range.
Note, that unlike the case of viscose fluid and resistive MHD,  {\it  the ratio of the linear dynamic to resistive terms is  scale-independent}  in EMHD.

On the other hand, the non-linear and resistive time scale become comparable at
\be
{L_{\rm res}  \over L} = {1 \over S_H} { B_0^2 \over \delta B_L^2 }
\ee
Thus, for  $\delta B_L  \sim B_0$ and $S_H \gg1$ there is a separation of cascade scales, and there is an inertial range.  This requires that the magnetic diffusivity satisfies
\be
\eta < \eta _{crit} = {c^2 \om_B \over \om_p^2}  {\delta B_L^2 \over B_0^2}
\ee

\subsection{Locality and strength of the quasi-isotropic EMHD Kolmogorov-like spectrum}
\label{locality}

In the thermodynamic equilibrium we expect $n_k \propto T/\om_k  \propto k^{-2}$, while in the EMHD cascade $n_k \propto T/\om_k  \propto k^{-3}$;  thus, the energy in the cascade will go to larger $k$ - whistlers obey a direct energy cascade.

Kolmogorov-type  cascades require locality, that the  interaction integral converges.  Assuming that $k_1 \ll k, k_2$, the interaction matrix of whistler waves  scales as
\be
V \propto k_1^{m_1} k^{m-m_1}, m_1=1, \, m=3
\ee
(For whistlers,  $V \propto (k_1 k_2/k)  \times (  k_1 $ or $ k_2)$). 
Infrared locality requires \citep{Zakharov}  $z_{IR}= 2 m_1 - m +1 >0$. In our case $z_{IR}=0$, thus, the infrared locality is weakly  violated (see also Ref. \cite{boldyrev95}). Since the  infrared locality is violated, this has important implication for transfer of energy to large scales and for the overall existence of the steady universal Kolmogorov-like spectrum.

Ultraviolet locality, in the limit $k_{1,2} \gg k$ of the 3-wave interaction requires ($\om \propto k^ \alpha$)
$z_{UV}=  2 m_1 - m + \alpha >0$. In our case $\alpha=2,\, m_1=1 $ or $m_1=2 $,  in the limit $k_{1,2} \gg k$, $V  \propto k_1^2 k_2^2 k^{-1}$, so that $z_{UV}=3> 0$: that the ultraviolet locality is satisfied.

We can also compare the non-linear time scale with the wave frequency
\be
{\nu_k \over \om_k } = {\delta B_k ^2 \over B_0^2} = {1\over k L} {\delta B_L ^2 \over B_0^2}  \leq 1 
\ee
Thus, the cascade becomes weaker for larger $k$. This is consistent with the general condition for  the weak cascade.
Comparing intensities of 3-wave and 4-waves interaction, 
for $\alpha >  (5-d)/3$ the  turbulence is weak for large $k$ \citep{Zakharov}. In our case $\alpha = 2> 2/3$, the turbulence is weaker at larger $k$. Thus, {\it  whistler cascade becomes weaker for large $k$}.  Since  these estimates come from comparing intensities of 3-wave and 4-waves interaction, they  are inapplicable to the \Alfven MHD turbulence.

Note, that the cascade  accelerates towards  small scales, becoming  explosive. A wave number of a perturbation with the initial $k \sim 1/L$ evolves according to
($\partial_ t k  = \nu_k k$, $\nu_k$ is given by Eq. (\ref{nuk})),
\be
k(t) =  \left( 1- { t \over t_H} {\delta B_L^2 \over B_0^2}  \right)^{-1} {1\over L}
\label{kex}
\ee
Thus, there is accelerating, explosive transfer on time scale 
\be
t_{casc} \sim t_H  { B_0^2 \over \delta B_L^2 }
\label{tcasc}
\ee
Though the cascade develops explosively, at each $k$ the non-linear interaction time remains smaller than the wave frequency, so the cascade, formally, remains weak. The fact that the cascade develops very fast at high $k$ makes the investigation of the time-asymptotic structure and the stability  very challenging.  We expect that at high wave numbers other physical effects, like resistivity and electron inertial would become important and resolve the formal singularity in Eq. (\ref{kex})

\subsection{Localized resistive heating}

For \Bf\  (\ref{deltaB}) the fluctuations of the current  increase with $k$: 
\be
\delta J = {4\pi \over c} \delta B_L \sqrt{ k \over L}
\ee
The resistive heating is then (in ergs cm$^{-3}$ s$^{-1}$)
\be 
\dot{\epsilon}= \eta J^2 = \eta \delta B_L^2 {k \over L}
\label{epsilon}
\ee
At the inner scale the dissipation rate is
\be 
\dot{\epsilon}_{\rm res} ={ \delta B_L^4  \over  B_0^2 t_H}= { \delta B_L^2 \over t_{casc}}
\ee
(so that all the cascade energy is dissipated on the cascade time scale). 

Localized resistive dissipation will be balanced by heat conduction. 
Since the thermal conduction flux $\propto  k $, the ratio of heating (Eq. (\ref{epsilon}))  to cooling  is independent  of $k$. In most applications  the thermal conductions times are much shorter than the Hall time, so that   the  localized dissipation by the EMHD cascade is likely not to be important.

\section{Conclusion}

In this paper we studied the  plasma dynamics and turbulence in the particular regime of inertialess electron MHD. Our main result is that 
analogies with the \Alfven turbulence cannot be transferred to the EMHD turbulence. First, there is no energy  principle in EMHD: an arbitrary EMHD configuration  does not first settle to a preferred state and then decay slowly, like an MHD plasma does. Instead, EMHD behaves similarly to the incompressible fluid, dissipating energy via the turbulent cascade.

 We considered in detail the three-wave interaction of whistlers and argued that it is very different from the MHD case: 
 (i) in EMHD {\it only harmonic}  whistlers are exact (non-linear) solutions; (ii)  co-linear whistlers do not interact;  (iii)  whistler modes with very different $k$ and $\theta$ can interact resonantly, via three-wave coupling, including those with zero frequency wave.Such processes are allowed both kinematically (the resonance condition is satisfied), as well dynamically (non-vanishing interaction matrix).
 We argue that such strong coupling between waves  propagating at different angles would in the general case eventually prevent a formation of highly anisotropic spectrum.

We have solved numerically the wave kinetic equation for various injection spectra. We found that the results of the simulations are highly dependent on the injection spectrum and on the numerical details. Often, 
as time progresses the spectrum becomes quasi-isotropic.
 On the other hand, {\it transiently} the cascade does develop highly anisotropic structures and suggest that a number of numerical simulations did catch this transient effect. Note that the fact that the cascade becomes very fast at wave numbers exceeding the injection scale, by a factor as high as $k^6$ (and, actually, accelerating) makes  it numerically challenging even for simple kinetic simulations that we adopted.

Thus, using analytical and numerical calculations we argue that the EMHD turbulence may  become quasi-isotropic, in a sense that it is not limited to a narrow range of angles, but may still behave broadly anisotropic. 
Based on this, and the  scaling of the interaction matrix, we find that the stationary  whistler turbulence should obey a $k^{-2}$ spectrums, as initially suggested in Ref.  \cite{RG}. On the other hand, the fact that the transfer rate becomes explosive and that the spectrum shows a IR divergence raises the questions about the stability and universality  of this  spectrum.

The key limitation of our approach is the assumption of constant underlying density. It is well known \citep{Kingsep,1994PhR...243..215G} that density gradients lead to whistler drifts that may lead to the formation of current sheets approximately in one Hall time. Formation of current sheets may have important implications for the behavior of the  \Bf. We will incorporate the density drifts in a future work.

I would like to thank Stanislav Boldyrev, Jungyeon Cho, Perter Goldreich, Alexander Lazarian,  and Christopher Thompson for numerous enlightening discussions.   This research was partially supported by NASA  grants NNX12AO85G,  NNX12AF92G  and by the Simons Foundation.

 \bibliographystyle{apsrev}
  \bibliography{/Users/maxim/Home/Research/BibTex} 
  
  \appendix
  
\section{Interaction of zero frequency mode and whistlers: parametric instability}
\label{parametric}

In a complementary approach, let us consider interaction of a single  transverse  zero frequency mode, with the wave vector orthogonal to the \Bf,  and an oblique  whistler waves. The background field is  $B_x(y,z)$;  such structure still satisfied the stationary condition $\nabla \times ( \nabla \times \B \times \B) =0$. For simplicity, let us assume 
$ B_x = B_0 + \delta B_0 \cos( k_0 y)$. Let us then impose whistler  perturbations $\delta \B =\{\delta B_x,\delta B_y, \delta B_z\} e^{- i ( \om t -k_x x)}$. Using the $\div \B=0$ condition and assuming small initial perturbations $ {\delta B_y / B_0} \ll 1$ , we find
\be 
{{\delta B_y}^{\prime \prime}  \over \delta B_y} = k_x^2 - { \om^2 \over B_0^2 k_x^2} - \left( k_0^2 - 2 {\om^2 \over B_0^2 k_x^2}\right) {\delta B_y \over B_0} \cos (k_0 y)
\ee
where prime denotes differentiation with respect to $y$. 
For small frequencies, $\om \ll B_0 k_x^2$, this  equation reduces to Mathieu's equation,
\be 
{{\delta B_y}^{\prime \prime} \over \delta B_y} = k_x^2 - k_0^2 {\delta B_y \over B_0} \cos (k_0 y)
\label{Mathieu}
\ee
Mathieu' s equation (\ref {Mathieu}) is well known in mathematical 
physics \citep {MCLACHLAN}. It has solutions which show instability 
bands centered (for small amplitude of  modulation $ {\delta B_y / B_0}\ll 1$)  on 
$  k_x^2= k_0^2  m/2$, where $m$ is integer. 
Thus, for a given periodic transverse variations with the wave number $k$, there are whistler modes that would produce exponentially growing \Bf\ (as functions of $y$).  Importantly, the effect is  {\it linear} in the amplitude of the whistler mode: regardless of the strength of the wave it 
 can lead to the creation of divergent \Bf\ structures. These structures are created on  scales of the order of the scale of the \Bf\ inhomogeneity. Temporarily, they will be created on a whistler time  scale. The fact that the parametric instability is linear in the amplitude of the wave and that it can operate in a constant density background distinguishes it from the non-linear Burger-type instability in the density gradient \citep{Kingsep}.

An important limitation of the above approach is that it  assumes that there exists  {\it a single} zero frequency mode. For the parametric instability it is required that the waves keep coherence over long temporal and/or spacial scales. Presence of other modes will destroy such coherence. 

  \end{document}